\documentclass[]{aastex63}

\RequirePackage{xspace}
\RequirePackage{amsmath}
\RequirePackage{amsfonts}
\RequirePackage{amssymb}

\def\ifundefined#1{\expandafter\ifx\csname#1\endcsname\relax}

\makeatletter
\newcommand*{\rom}[1]{\expandafter\@slowromancap\romannumeral #1@}
\makeatother

\def\la{\mathrel{\hbox{\rlap{\hbox{\lower4pt\hbox{$\sim$}}}\hbox{$<$}}}}
\def\ga{\mathrel{\hbox{\rlap{\hbox{\lower4pt\hbox{$\sim$}}}\hbox{$>$}}}}

\newcommand{\be}{\begin{equation}}
\newcommand{\ee}{\end{equation}}

\newcommand{\bea}{\begin{eqnarray}}
\newcommand{\eea}{\end{eqnarray}}

\ifundefined{ensuremath}\def\ensuremath#1{\relax\ifmmode{#1}}
\else${#1}$\fi\else\relax\fi
\ifundefined{nuc}\def\nuc#1#2{\relax\ifmmode{}^{#1}{\protect\text{#2}}
\else${}^{#1}$#2\fi}\else\relax\fi

\ifundefined{ion}
\newcommand\ion[2]{#1$\;${\ifx\@currsize\normalsize\small \else
\ifx\@currsize\small\footnotesize \else
\ifx\@currsize\footnotesize\scriptsize \else
\ifx\@currsize\scriptsize\tiny \else
\ifx\@currsize\large\normalsize \else
\ifx\@currsize\Large\large
\fi\fi\fi\fi\fi\fi
\rmfamily\rom{#2}}\relax}\else\relax\fi

\newcommand{\kmps}{\ensuremath{\text{km}~\text{s}^{-1}}\xspace}

\newcommand{\nni}{\ensuremath{\nuc{56}{Ni}}\xspace}

\def\tstd{\ensuremath{\tau_{\text{std}}}\xspace}

\newcommand{\phx}{\texttt{PHOENIX}\xspace}
\newcommand\phxO{\texttt{PHOENIX/1D}\xspace}

\makeatletter

\newcommand\Autoref[1]{\@first@ref#1,@}
\def\@throw@dot#1.#2@{#1}\def\@set@refname#1{    \edef\@tmp{\getrefbykeydefault{#1}{anchor}{}}    \def\@refname{\@nameuse{\expandafter\@throw@dot\@tmp.@autorefname}s}}
\def\@first@ref#1,#2{  \ifx#2@\autoref{#1}\let\@nextref\@gobble  \else    \@set@refname{#1}    \@refname~\ref{#1}    \let\@nextref\@next@ref  \fi  \@nextref#2}
\def\@next@ref#1,#2{   \ifx#2@ and~\ref{#1}\let\@nextref\@gobble   \else, \ref{#1}   \fi   \@nextref#2}

\makeatother

\usepackage[capitalize]{cleveref}

\graphicspath{{./figs/}}

\begin{document}

\title{Ultraviolet Line Identifications and Spectral Formation Near Max-Light 
in Type Ia Supernovae 2011fe}

\author[0000-0002-7566-6080]{James M. DerKacy}
\affiliation{Homer L. Dodge Department of Physics and Astronomy, University of 
Oklahoma}

\author[0000-0001-5393-1608]{E. Baron}
\affiliation{Homer L. Dodge Department of Physics and Astronomy, University of 
Oklahoma}
\affiliation{Hamburger Sternwarte, Gojenbergsweg 112, 21029 Hamburg, Germany}

\author{David Branch}
\affiliation{Homer L. Dodge Department of Physics and Astronomy, University of 
Oklahoma}

\author[0000-0002-4338-6586]{Peter Hoeflich}
\affiliation{Department of Physics, Florida State University}

\author{Peter Hauschildt}
\affiliation{Hamburger Sternwarte, Gojenbergsweg 112, 21029 Hamburg, Germany}

\author[0000-0001-6272-5507]{Peter J. Brown}
\affiliation{George P. and Cynthia Woods Mitchell Institute for
  Fundamental Physics \& Astronomy, Mitchell Physics Building, Texas
  A. \& M. University, USA}  

\author{Lifan Wang}
\affiliation{George P. and Cynthia Woods Mitchell Institute for
  Fundamental Physics \& Astronomy, Mitchell Physics Building, Texas
  A. \& M. University, USA}  

\submitjournal{\apj}
\received{\today}
\revised{\today}
\accepted{\today}
\correspondingauthor{James M DerKacy}
\email{jmderkacy@ou.edu}

\begin{abstract}
We present ultraviolet line identifications of near maximum-light HST 
observations of SN 2011fe using synthetic spectra generated from both SYNOW and 
\texttt{PHOENIX}. We find the spectrum to be dominated by blends of iron group 
elements Fe, Co, and Ni (as expected due to heavy line blanketing by these 
elements in the UV) and for the first time identify lines from C~IV and Si~IV in
a supernova spectrum. We also find that classical delayed detonation models of Type Ia supernovae are 
able to accurately reproduce the flux levels of SN 2011fe in the UV. Further analysis
reveals that photionization edges play an important role in feature formation in the
far-UV, and that temperature variations in the outer layers of the ejecta significantly
alter the Fe~III/Fe~II ratio producing large flux changes in the far-UV and velocity
shifts in mid-UV features. SN 2011fe is the best observed core-normal SNe Ia, 
therefore analysis its of UV spectra shows the power of UV spectra in discriminating between different 
metalicities and progenitor scenarios of Type Ia supernovae, due to the fact 
that the UV probes the outermost layers of the Type Ia supernova, which are most 
sensitive to metalicity and progenitor variations.
\end{abstract}

\section{Introduction}

The UV spectrum of Type Ia supernovae (SNe Ia) is important for
understanding the nature of the explosion, since it both forms
throughout the supernova atmosphere \citep{bongard2008} and probes the
outermost layers \citep{lentz2000,H17}. Variations in the UV spectra
with redshift have been observed
\citep{ellis08,foley2012,maguire_uv_12}. \citet{foley_kirshner} used a comparison of the
``twins'' SN 2011by and SN 2011fe to deduce a variation in progenitor
metalicity of the two supernovae, using the models of
\citet{lentz2000} to infer that progenitors of SN 2011by and SN
2011fe were supersolar and subsolar, respectively. \citet{Brown2015}
found that they could reproduce the same results as those of
\citet{foley_kirshner} using only photometry obtained by the Neil Gehrels 
Swift Observatory, hereafter \textit{Swift}. However, they
found that the UV flux levels of the \citet{lentz2000}
models were far too high, due to the structure of W7 model
\citep{nomw7}. Recently, \citet{Pan2020} used grism
spectroscopy of a sample of SNe Ia obtained with \textit{Swift} and correlated
it with the progenitor metalicity. Using \textit{Swift} photometry, this result
has been challenged \citep{Brown2019}.

Several studies have identified UV features in Type Ia SNe using
a variety of models and methods, but few line identifications are consistent across
these works (\cref{table:prev_lineid} lists the identified 
features from the literature). \citet{branch1986} used an early version of the
parameterized SYNOW code to identify features in the near-UV of an IUE
spectrum of SN 1981B at -2 days relative to maximum
light. \citet{kirshner_92A} analyzed the first high-quality
near-maximum light spectrum of a Type Ia; a combined IUE and HST
spectrum of SN 1992A at +5 days. Using a parameterized synthetic LTE
spectrum of a delayed-detonation explosion model, they were able to
identify many of the mid-UV features, although some identifications
are described as tentative. \citet{hachinger2013} studied the HST
spectrum of SN 2010jn at -0.3 days. Using an updated version of the
Monte Carlo spectrum-synthesis code of
\citet{mazzali_lucy}, they calculated synthetic spectra using the  density
profiles of the W7 model \citep{nomw7} and the WDD3 delayed-detonation model from \citet{iwamoto99}
with abundances determined via abundance tomography. 
\citet{mazzali14} (hereafter M14) performed a similar analysis to 
Hachinger using the WDD1 delayed-detonation model of \citet{iwamoto99} 
on a time series of HST spectra of SN 2011fe 
covering from -13.1 to +40.8 days, including the first spectrum to 
have significant coverage of the far-UV (+3.4 days).

Other factors influencing the UV spectra of Type Ia SNe, such as density
and model luminosity, have also been investigated. Using much of the same
methodology as \citet{hachinger2013} and M14, \citet{sauer2008} altered the
power law index of the density profile in the outer layers of the W7 model,
concluding that steeper density profiles result in more UV flux and 
a better match to observed UV spectra than those with shallower
density profiles; all without producing large changes in the optical  
spectra. \citet{walker2012} varied the luminosity of their models, finding that
high luminosity models produce more UV flux, but more featureless UV spectra.
The interpretation of these results is made more complex since they
simultaneously 
changed the density structure of their models when varying the luminosity.

We present synthetic spectra from SYNOW and \texttt{PHOENIX} 
which are used to identify all major features from the far-UV to the near-UV 
in the +3.4 day HST spectrum of SN 2011fe first presented in \citet{mazzali14}. 
A suite of \texttt{PHOENIX} spectra are then used to further examine the 
impact of different mechanisms that combine to form the UV spectrum, and 
to determine the temperature dependence of multiple features, 
which may be useful in constraining physical parameters within the ejecta. 
Section \ref{sec:codes} outlines the models and spectral synthesis codes 
used in this work, with the line identifications from these spectra presented 
in Section \ref{sec:ids}. Section \ref{sec:spec_form} further examines line 
formation mechanisms that play an important role in the UV. Section 
\ref{sec:discussion} places these results into the broader context of work on 
UV spectra of Type Ia SNe. Section \ref{sec:conclusion} summarizes our conclusions. 

\section{Spectral Modeling} \label{sec:codes}
This work makes use of two different spectral synthesis codes, SYNOW and 
\texttt{PHOENIX}, capable of making line identifications in supernova spectra. 
Both codes rely on different assumptions and underlying physics to synthesize 
the spectra, providing a useful check on the other and improving our confidence 
in each line identification. The two codes are briefly summarized below.

\subsection{SYNOW}

SYNOW is designed to simulate supernova spectra and relies on simple 
assumptions that describe the supernova during the photospheric phase, 
including: spherical symmetry, homologous expansion ($v \propto r$), a sharp 
photosphere that emits a blackbody continuum, and lines formed via resonance 
scattering, which are treated in the Sobolev approximation. SYNOW does not perform 
continuum transport; nor does it calculate ionization ratios or solve rate 
equations. Its primary purpose is to account for multiple line scattering 
so that it can be used in the empirical spirit to make line identifications, 
estimate the photospheric (or pseudo-photospheric) velocity, and roughly 
determine the velocity interval within which each ion is detected. The synthetic 
supernova spectrum generated by SYNOW consists of blended P-Cygni profiles 
(consisting of an unshifted emission component with a blueshifted absorption 
component) superimposed on the blackbody continuum. 

For each ion included in the fit, the optical depth of a reference line at one 
velocity (typically the photospheric velocity) is a fit parameter, and the 
optical depths of the other lines of the ion at that velocity are calculated 
assuming a Boltzmann excitation temperature $T_{exc}$. Typically, the strongest
optical line of an ion is chosen as the reference line. To limit the parameter 
space of the fit, $T_{exc}$ is chosen to have the same value for each ion, 8000 
K for the fit shown here. All line optical depths decrease exponentially with 
velocity, according to $\tau(v) = \tau(v_0) e^{-(v-v_0)/v_e}$, where the 
e-folding velocity $v_e$ is generally taken to be 1000 km s$^{-1}$. 
Therefore, the important parameters 
of the fit are the photospheric velocity $v_{0}$, the optical depths of the 
ion reference lines, the velocity extent of each ion, and the e-folding 
velocity $v_e$ of each ion. The fit is optimized by eye, as is standard within 
the community. More information on SYNOW can be found in 
\citet{jeffery_branch_1990} and \citet{branch2005,branch2006}.

\subsection{\texttt{PHOENIX}}
 
\begin{figure}
\centering
\plotone{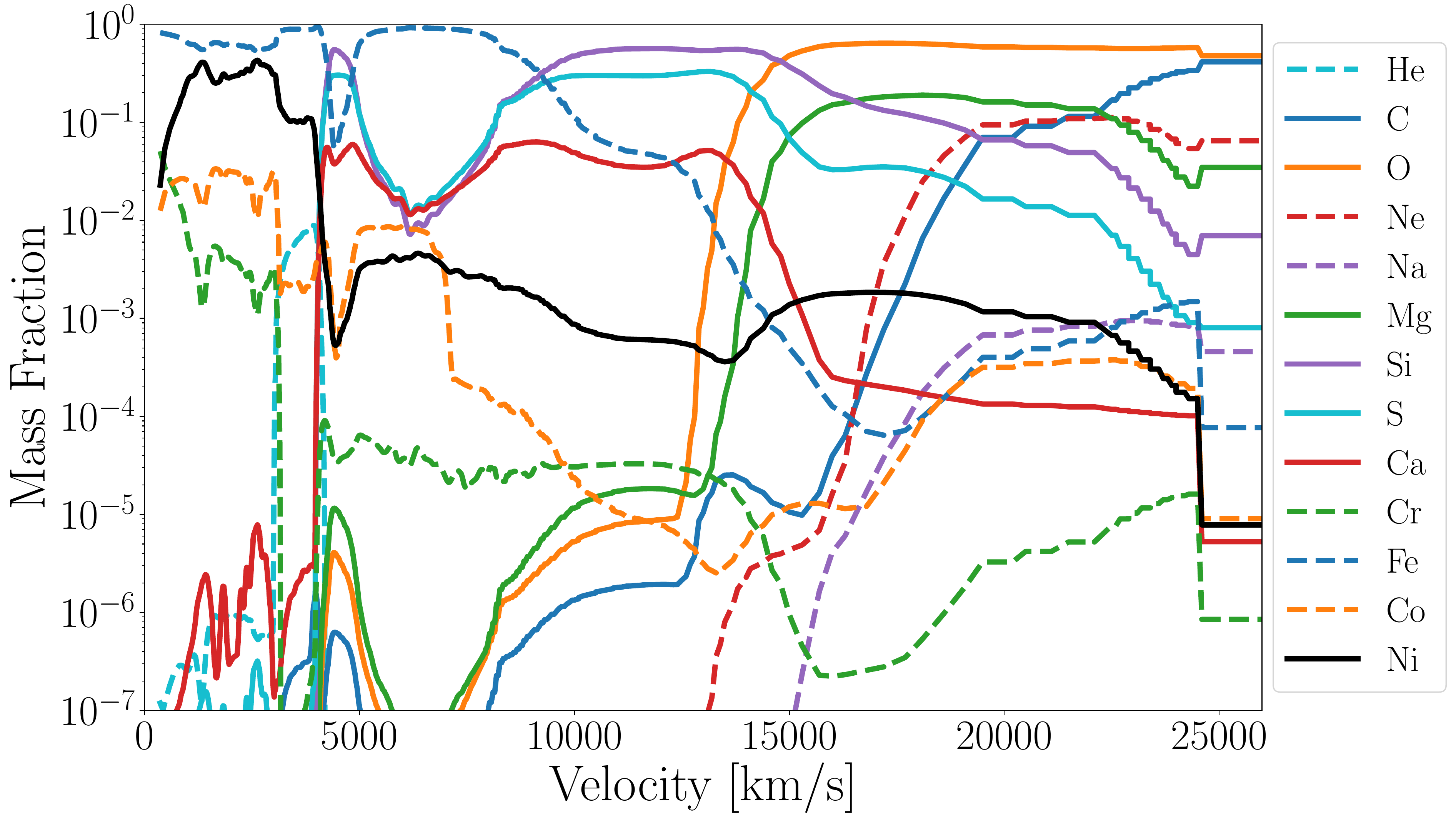}
\caption{Final distribution of elements in the DD model as a function of velocity.
All abundances are held constant above 25,000 km s$^{-1}$.}
\label{fig:mod_struc}
\end{figure}

This work makes use of a generalized 1-D delayed detonation (DD) model
first presented in \citet{dominguez2001} which reproduces the
light curves and spectra of Branch-normal supernovae and was previously
shown to well match the pre-maximum light spectra of SN 2011fe in
\citet{baron2015} (specifically, we use the prompt DDT model whose
density structure is shown in their Figure 1). The model starts with a C/O white dwarf taken from
the core of an evolved 5M$_\odot$ main-sequence star. This core
approaches the Chandrasekhar mass through accretion, and an explosion
is spontaneously triggered when the central density reaches $2.0\times
10^9$ g cm$^{-3}$. The deflagration-to-detonation transition occurs at
a density of $2.3 \times 10^7$ g cm$^{-3}$. The metalicity of the model is 
$Z_{\odot}/20$ where the metalicity  is defined as the ratio of the
iron abundance to the solar iron abundance. The abundance structure of the
model is shown in \cref{fig:mod_struc}. Here, we recalculate the 
spectra using \texttt{PHOENIX} at 23 days after explosion, corresponding 
to the +3.4 day HST spectrum of SN 2011fe first presented in M14. 

\phxO version \texttt{18} is a general purpose stellar atmospheres program. It 
solves the radiative transfer equation along characteristic rays in spherical 
symmetry including all special relativistic effects. \phxO solves the non-LTE (NLTE) rate 
equations. The effects of 
ionization due to non-thermal electrons produced from the $\gamma$-rays produced 
from the radioactive decay of $^{56}$Ni synthesized during the supernova 
explosion are included. The ions treated in NLTE are He I-II, C~I-IV, O~I-III, 
Ne I, Na I-II, Mg~I-III, Si~I-IV, S I-III, Ca II, Ti II, Cr~I-III, Mn I-III, 
Fe~I-III, Co~I-III, and Ni~I-III. These should encompass all the ions
that have features that significantly form 
the observed spectral features in SNe Ia.

Each model atom includes primary NLTE transitions, which are used to calculate 
the level populations and opacity, however, all the opacity is
included even if the levels of those lines are not in the model
atom. These weaker lines are treated in LTE using the true NLTE
occupation of the ground state.
This opacity implicitly affects the rate equations via their 
effect on the solution to the transport equation \citep{phx1999} and
ensures that no line transition is excluded. For ions not treated in NLTE
the  line opacities 
are treated with the equivalent two-level atom source function, 
using a thermalization parameter, $\alpha =0.10$ \citep{baron1996}. The 
atmospheres are iterated to energy balance in the comoving frame; while we 
neglect the explicit effects of time dependence in the radiation transport 
equations, we do implicitly include these effects, via explicitly including $P 
dV$ work and the rate of gamma-ray deposition in the generalized equation of 
radiative equilibrium and in the rate equations for the NLTE populations.

The outer boundary condition is the total bolometric luminosity in the 
observer's frame, and is the main tunable parameter in the simulations. 
The inner boundary condition is that the flux at the innermost zone 
(here $v = 700$~\kmps) is given by the diffusion equation. Converged 
models require 256 optical depth points to correctly obtain the Si~II 
$\lambda$6355 profile. The model is simulated at several different target
luminosities, after which the spectra are examined and the best fit determined,
again via``chi-by-eye".

\citet{baron2015} use \phxO version \texttt{16}, however the changes from 
version \texttt{16} to version \texttt{18} are confined to the 
\texttt{PHOENIX}/3D mode, updates to the molecular line lists (not included in 
either the previous or current calculations) and various bug fixes.  

\section{Line Identifications} \label{sec:ids}

Guided by the previous line identifications in near maximum-light spectra of 
Type Ia SNe outlined in \cref{table:prev_lineid}, we generate synthetic spectra 
to fit the +3.4 day HST observations of SN 2011fe. Our goal is to provide a 
complete set of line identifications for the spectral features in the near 
max-light UV spectra of Type Ia SNe, and in particular to address the 
disagreement over whether singly or doubly ionized iron group elements (IGEs) 
like Fe and Co are responsible for UV feature formation.

\begin{deluxetable*}{ccccc}
\tablecaption{Previous UV Line Identifications \label{table:prev_lineid}}
\tablehead{\colhead{$\lambda$ (\AA)} & \colhead{1981B (-2 d)\tablenotemark{a}} & 
\colhead{1992A (+5/6 d)\tablenotemark{b}} & \colhead{2010jn 
(+4.8 d)\tablenotemark{c}} & \colhead{2011fe (+0.1/+3.4 d)\tablenotemark{d}}}
\startdata
3300 & Co~II & - & Co~II, Co~III, Fe~III & Co~III, Fe~III \\
3090 & Fe~II & - & - & Si~III, Co~III, Fe~III \\
3010 & - & Fe~II, Co~II, Si~III & - & Si~III, Co~III, Fe~III \\
2820 & - & \textit{Mg~II}, Fe~II & Fe~II, Mg~II, Fe~III & Co~III, Fe~III \\
2650 & - & Mg~II, Fe~II & - & Mg~II, Fe~II \\
2470 & - & Fe~II & - & Fe~II, Co~II \\
2250 & - & Fe~II & - & Fe~II, Co~II, Ni~II \\
1950 & - & \textit{Cr~II} & - & - \\
1580 & - & - & - & - \\
1430 & - & - & - & Si~II, Co~II, Fe~III \\
1290 & - & - & - & - \\
\enddata
\tablecomments{Line identifications that are described as tentative/weak by the 
original authors are italicized.}
\tablerefs{$^{\textit{a}}$\citet{branch1986}; 
$^{\textit{b}}$\citet{kirshner_92A}; $^{\textit{c}}$\citet{hachinger2013}; 
$^{\textit{d}}$\citet{mazzali14}}
\end{deluxetable*}

\subsection{SYNOW Line ID's}
\begin{figure*}[ht]
\centering
\plotone{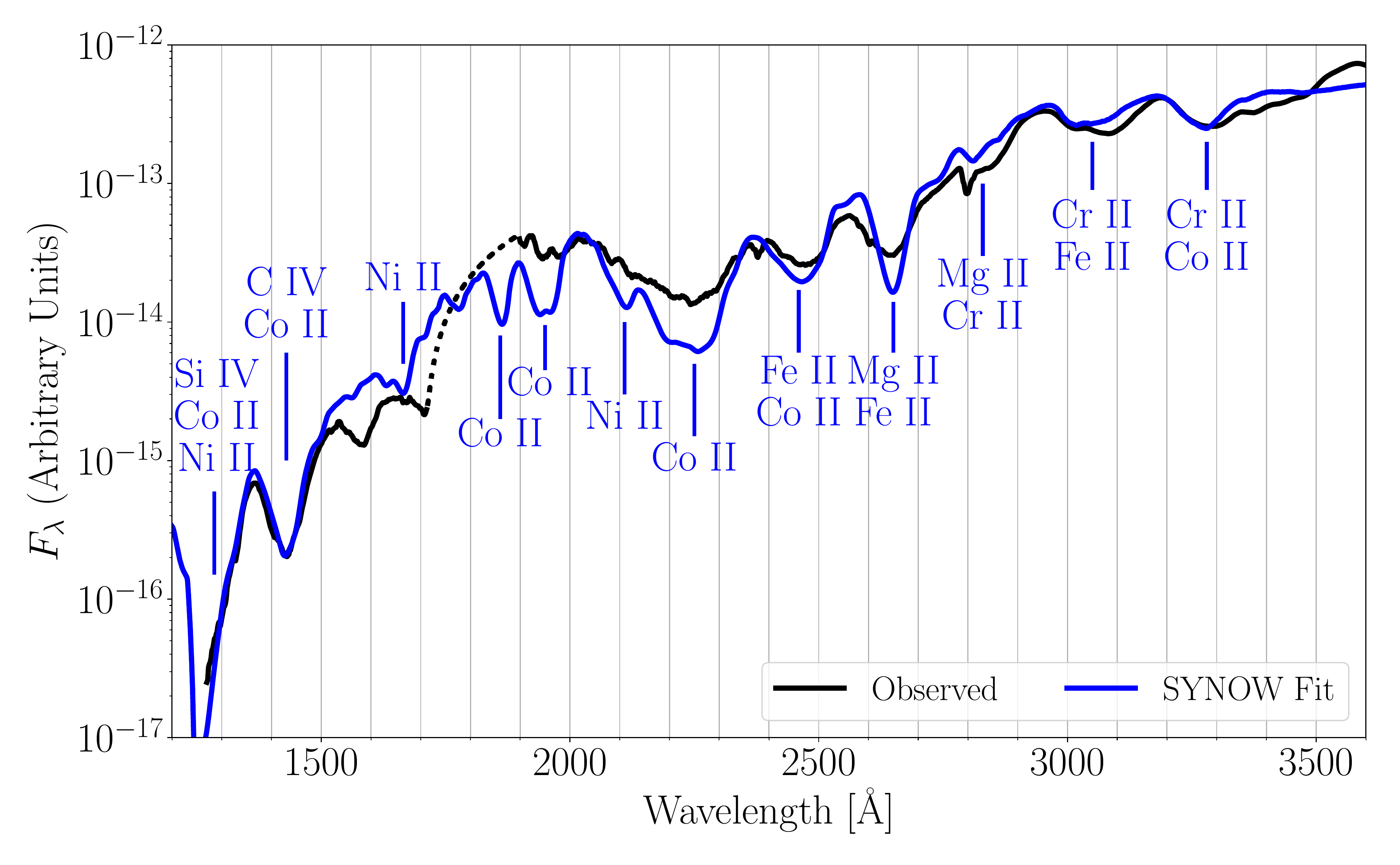}
\caption{SYNOW Fit with line identifications (in blue) as compared to HST 
observations (black). The gap in the spectrum from $\sim 1700$~\AA\ - 1900~\AA\
(dotted black line) is due to a gap in the wavelength coverage of 
HST. As such, any line identifications within or bordering this region should be 
considered tentative.}
\label{fig:branch_id}
\end{figure*}

\cref{fig:branch_id} shows the SYNOW spectrum and corresponding line 
identifications. Overall the SYNOW spectrum fits the observations well, although 
the lines are too strong in some mid-UV features. Compared to previous 
line identifications in near-max light UV spectra of Type Ia SNe (see 
\cref{table:prev_lineid}), the identifications of UV features differ 
significantly in the both the near and far-UV. For the first time, Cr~II is identified in the near 
max-light UV spectrum of a Type Ia, and the features centered around $\sim 
3300$~\AA\, $\sim 3090/3010$~\AA\, and $\sim 2820$~\AA\ features.\footnote{M14 
identify Cr~II in both the near-UV features at early times, but 
argue that these features change to blends of Fe~III and Co~III around -7 days} 
Additionally, the contributions from Fe and Co throughout the UV are attributed 
solely to the singly ionized states, not partially or wholly to the double 
ionized states as  found previously. In the mid-UV, we find better agreement 
with previous work, where the prominent features are identified as blends 
primarily of Fe~II, Co~II, and Ni~II, with the $\sim 2820$~\AA\ and $\sim 
2650$~\AA\ features requiring contributions from Mg~II lines. In the 
far-UV, the main features are caused by the strong resonance lines of C~IV and Si~IV
blended with weaker Co~II and Ni~II lines as the primary iron group element contributors. 

\subsection{\texttt{PHOENIX} Line ID's}

Previously, line identifications made with \texttt{PHOENIX} were determined 
via ``single-ion spectra'', where the converged model has all line
opacities artificially set to zero, except for the ion of interest
\citep{bongard2008}. However, in the UV line blanketing from IGEs and
blending of strong lines of intermediate mass elements (IMEs), 
unburned material, and IGEs in the UV contribute to 
the formation of the broad UV spectral features making the use 
of single-ion spectra difficult. Instead, we adopt the inverse approach, 
setting the line opacities in our ion of interest to zero and looking for 
changes in the flux near the spectral features by subtracting the spectra 
without the ion of interest from the full fit. While spectra are not additive 
in this manner, this approach provides us with the 
minimal contribution of the strong lines for each ion of interest to
the spectrum.  Another relatively minor complication is that
\texttt{PHOENIX} iterates the 
scattering problem when generating spectra so the mean intensity $J$,
is readjusted for the new (diagnostic) opacity and source
function. Since the opacity changes less in the inverse single-ion
approach than in the single-ion approach, the inverse approach should
be preferable for this particular effect.
The relative line strengths of ``inverse single-ion spectra" for the ion of 
interest will be underestimated in blended features due to photons scattering 
into the other lines comprising the blend. Locally normalized spectra 
\citep{jeffery2007} are used in the line identification process to account 
for the flux change of over four orders of magnitude from the near to far-UV. 
We only identify lines with a residual greater than 0.1 in order 
to avoid identifying weak lines that contribute to the line blanketing in the 
region but not the feature itself. While this choice of residual value is
somewhat arbitrary, residuals less than 0.1 are hard to distinguish when 
visually comparing spectra with and without the ion of interest, and typically
do not change the line shapes when excluded from the best fit spectrum (see the
weak Cr~II lines within \cref{fig:ige_res} for example).

\begin{figure*}[ht]
\centering
\plotone{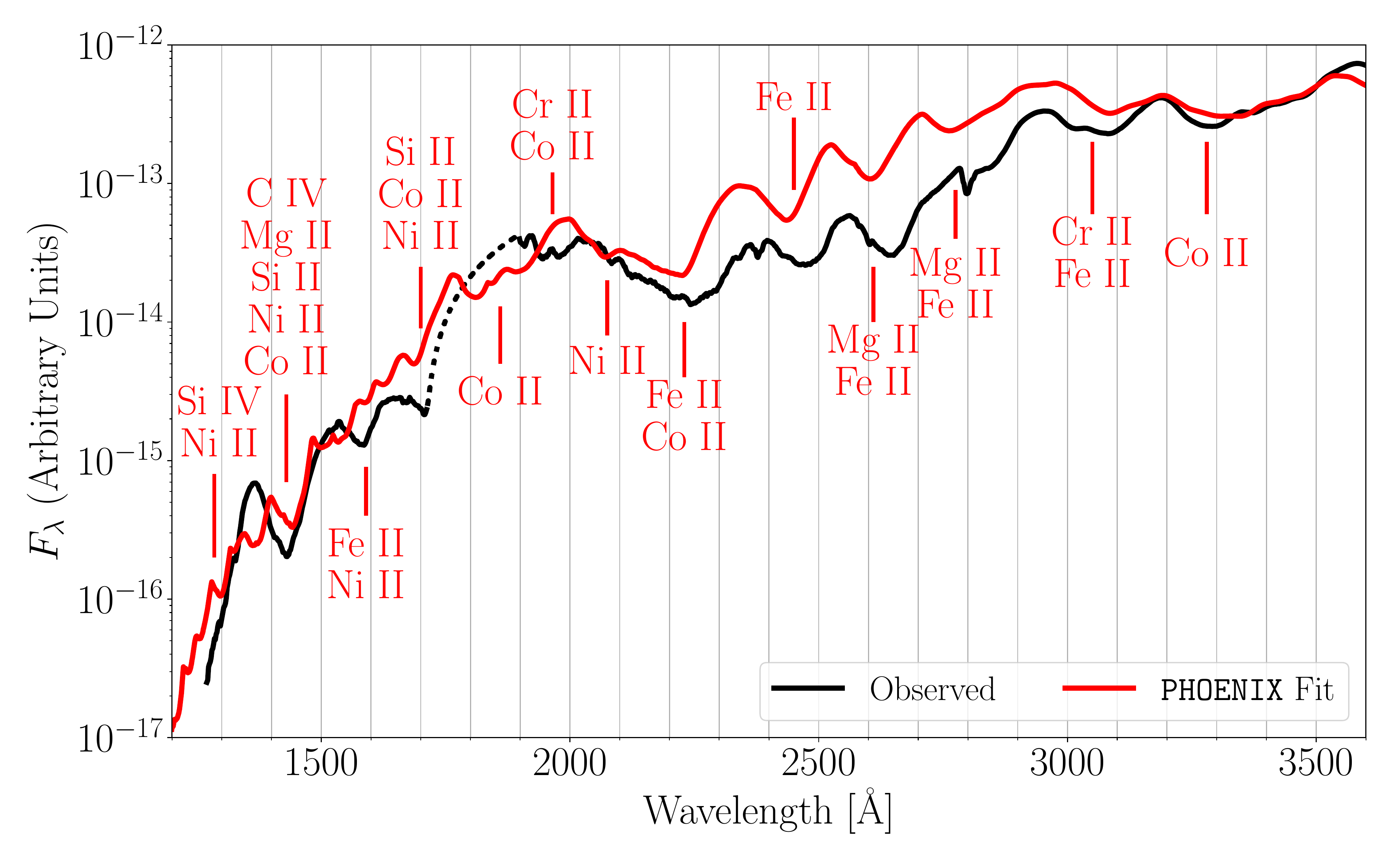}
\caption{Same as \cref{fig:branch_id}, but with the \texttt{PHOENIX} best fit 
and line identifications (in red). Features in the far-UV are noticeably more 
complex than those in the SYNOW fit, and include more ions making significant 
contributions to the line blending.}
\label{fig:phx_id}
\end{figure*}

\begin{figure*}
\centering
\gridline{\fig{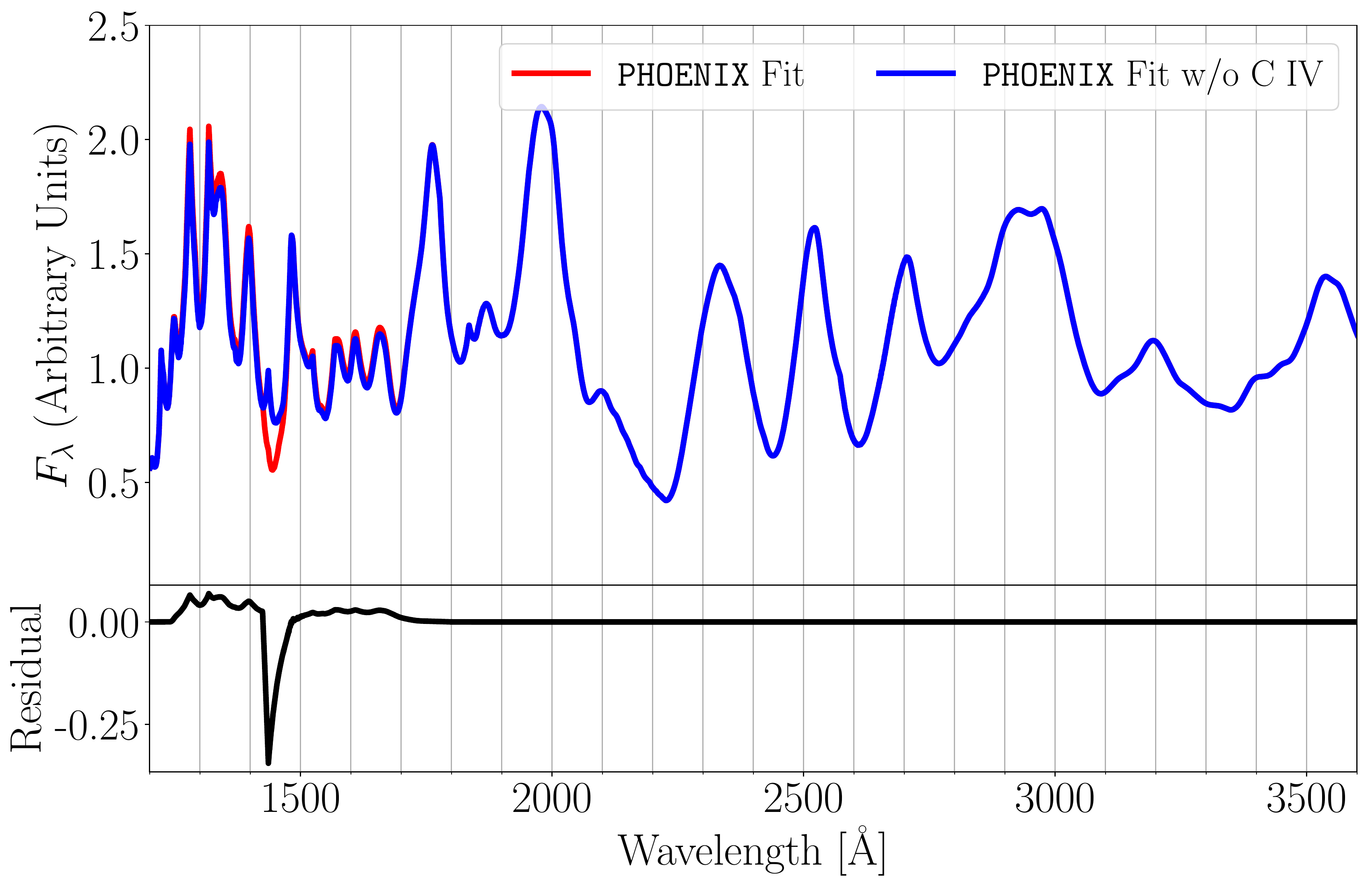}{0.45\textwidth}{(a) C IV}
		  \fig{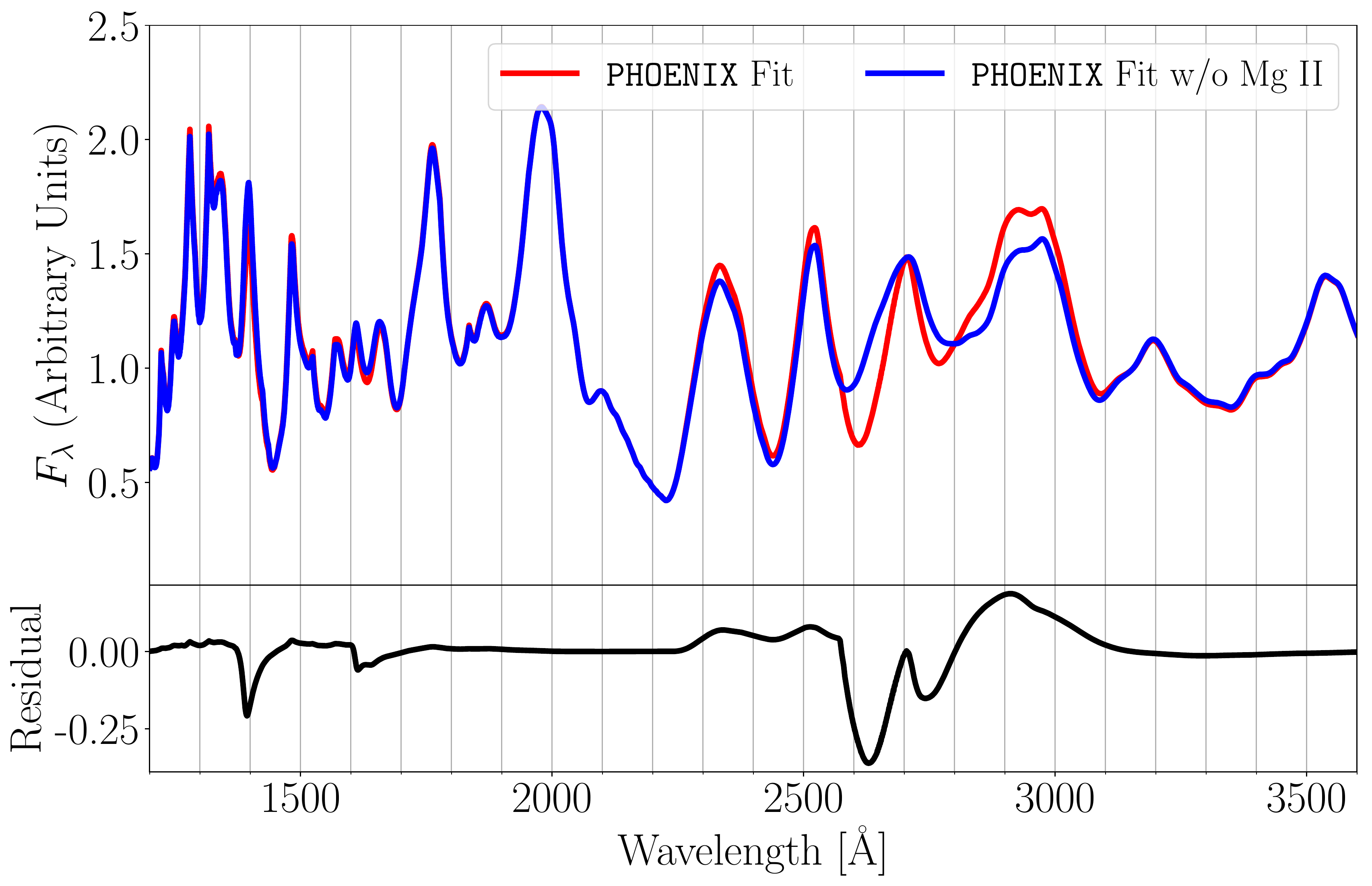}{0.45\textwidth}{(b) Mg II}}
\gridline{\fig{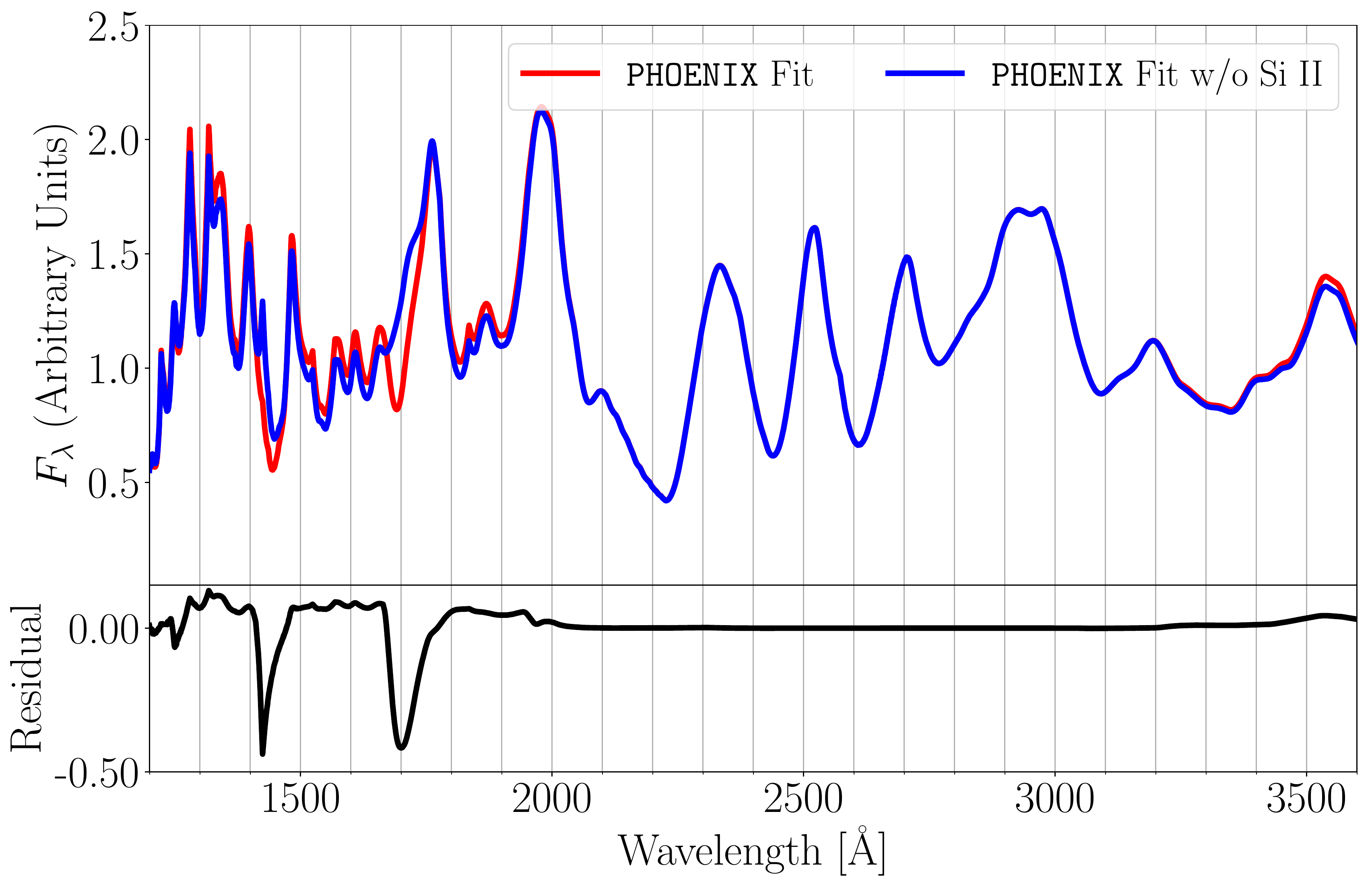}{0.45\textwidth}{(c) Si II}
		  \fig{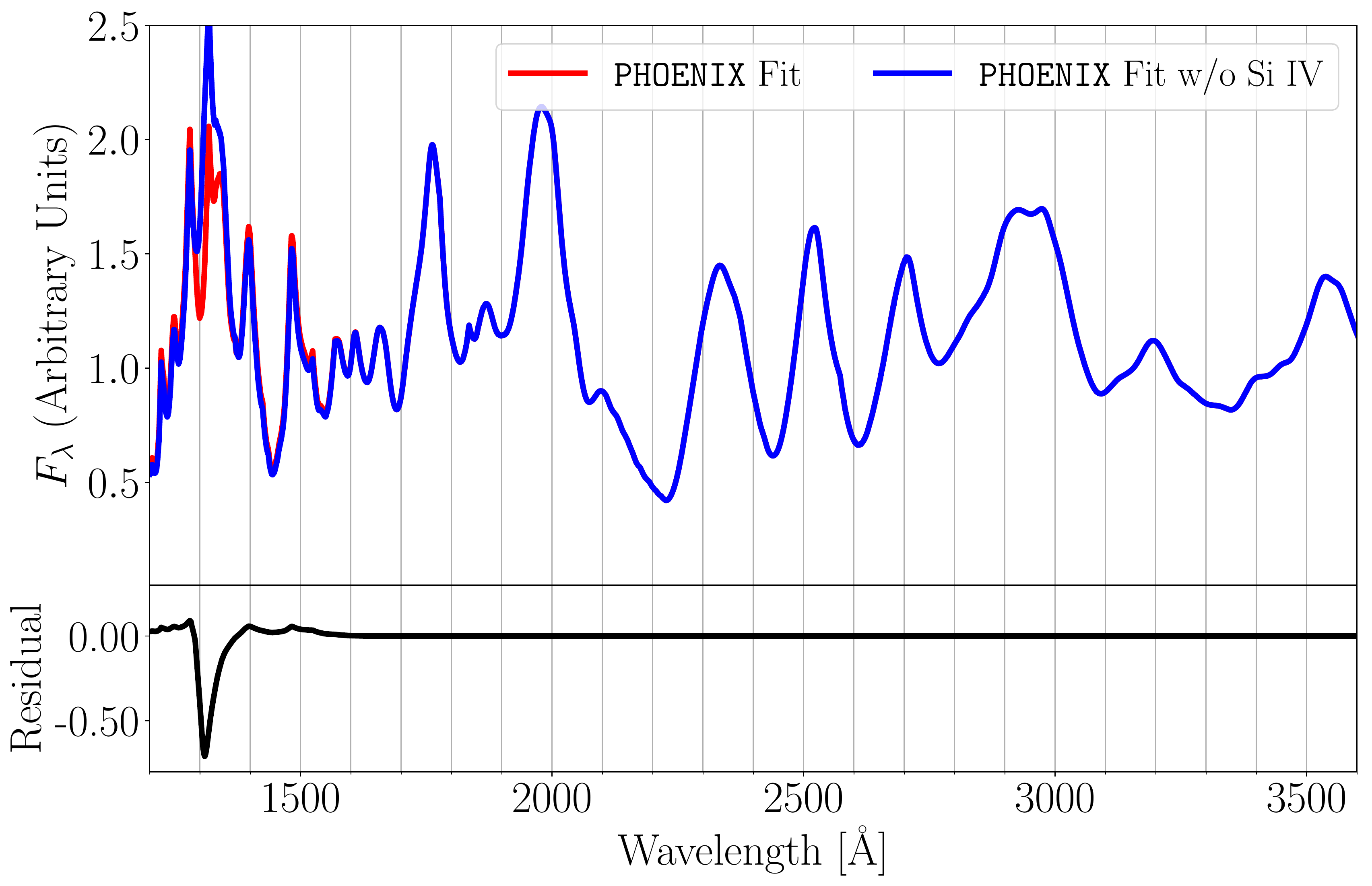}{0.45\textwidth}{(d) Si IV}}
\caption{Residual plots of unburned material and intermediate mass elements that 
contribute to feature formation in the UV. The top panel of each plot shows the 
locally normalized UV spectra of the full \texttt{PHOENIX} fit (in red) over 
plotted with the \texttt{PHOENIX} fit without the ion of interest (blue).}
\label{fig:ime_res}
\end{figure*}

\begin{figure*}
\centering
\gridline{\fig{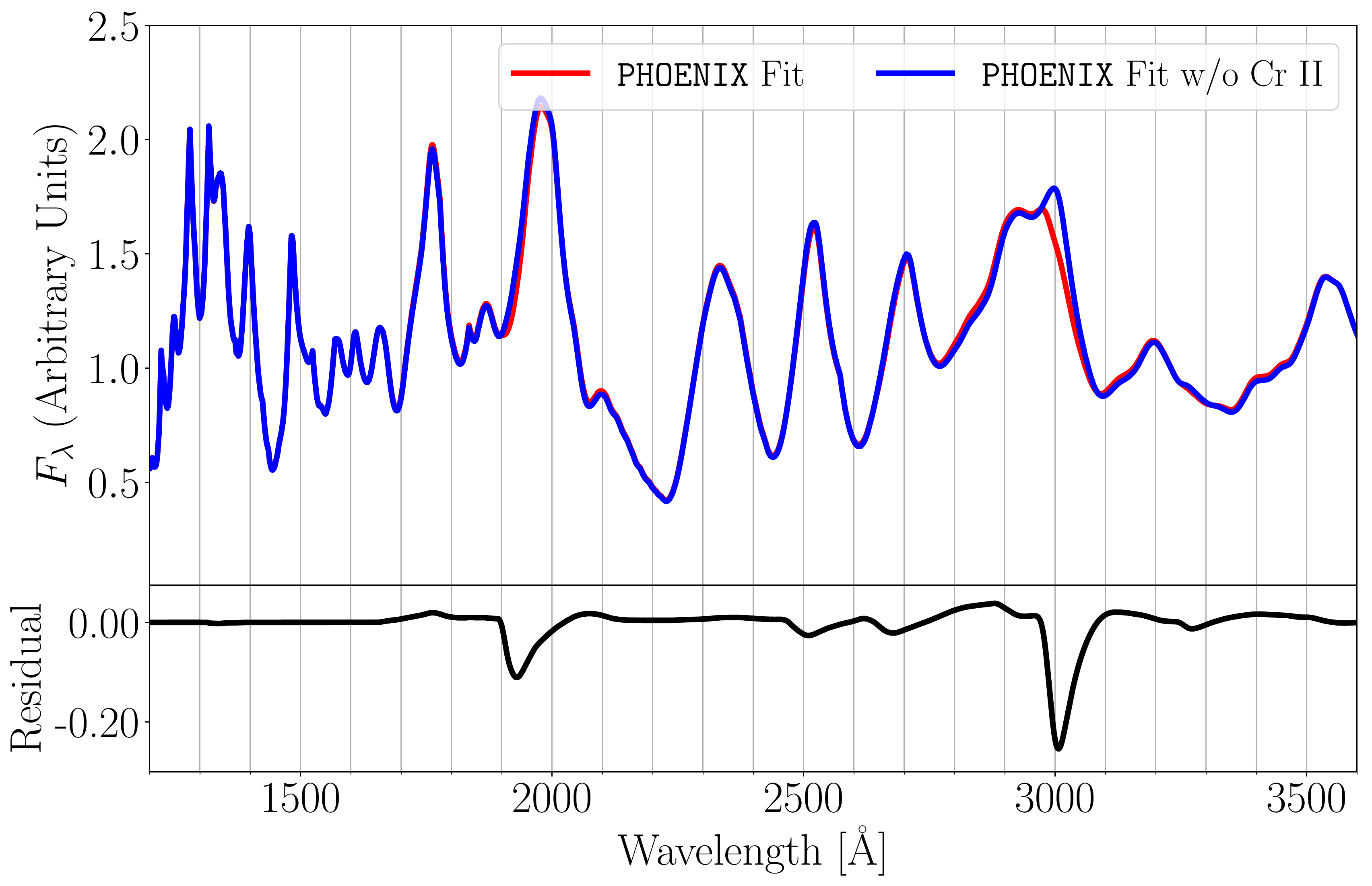}{0.45\textwidth}{(a) Cr II}
		  \fig{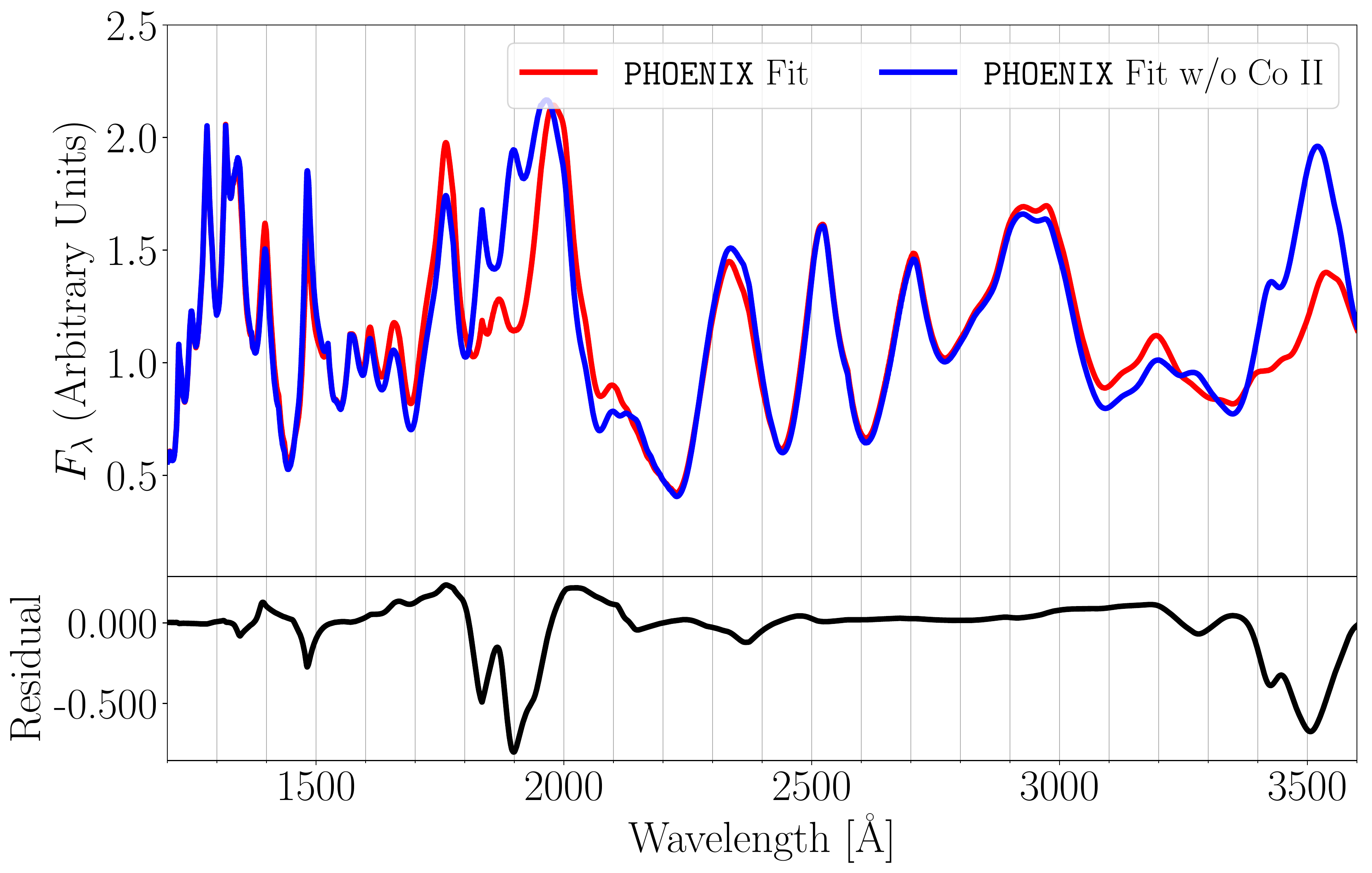}{0.45\textwidth}{(b) Co II}}
\gridline{\fig{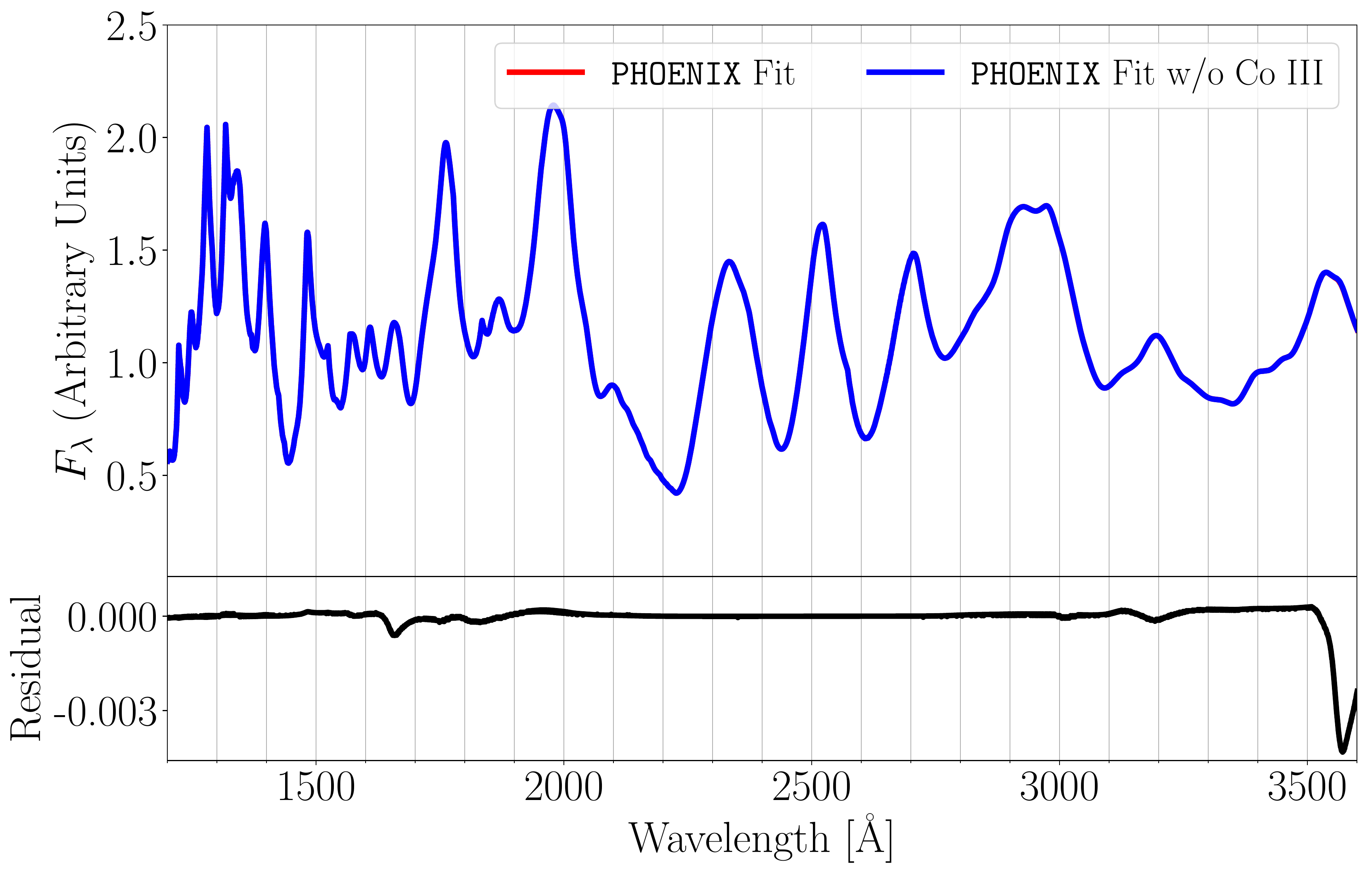}{0.45\textwidth}{(c) Co III}
		  \fig{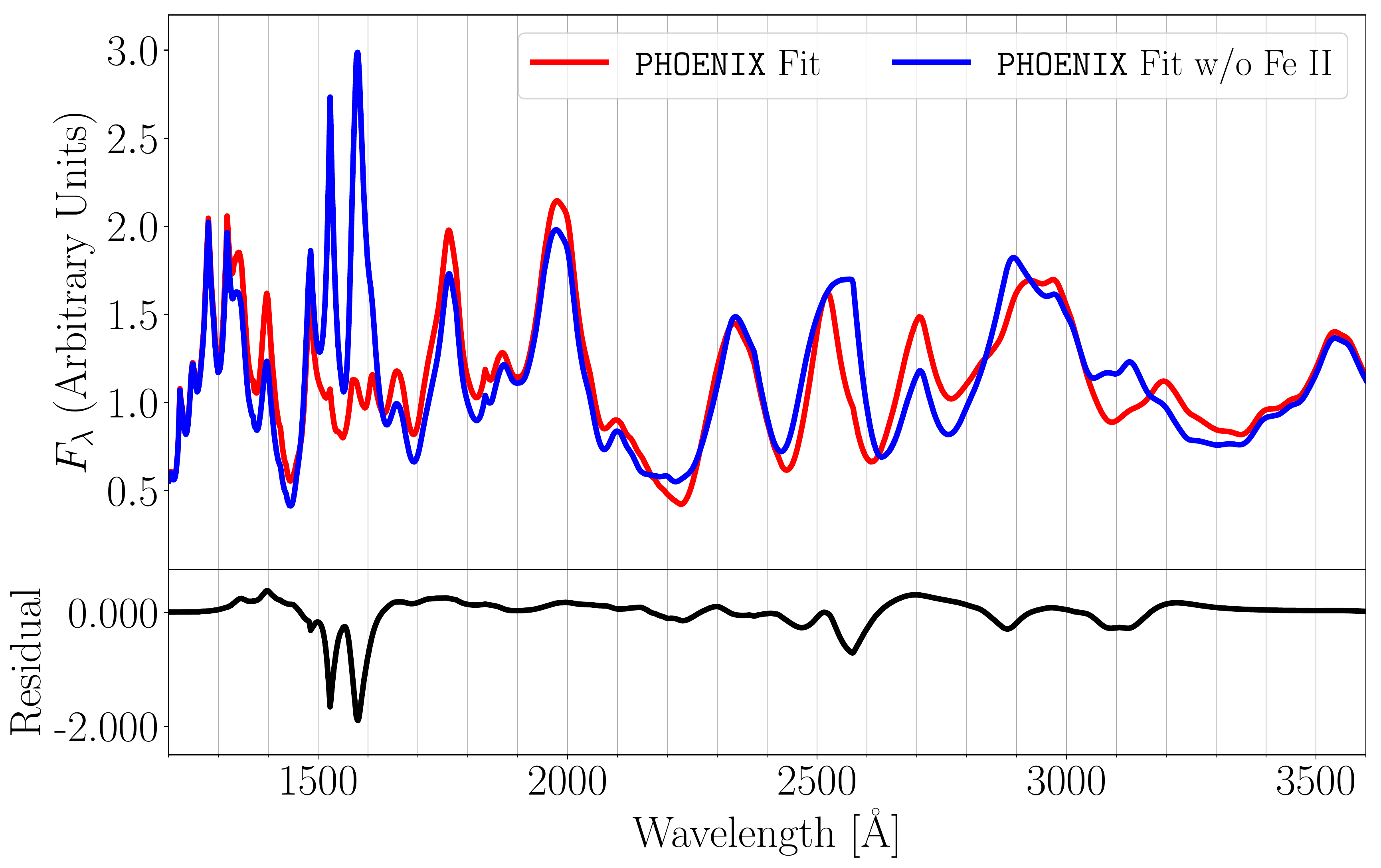}{0.45\textwidth}{(d) Fe II}}
\gridline{\fig{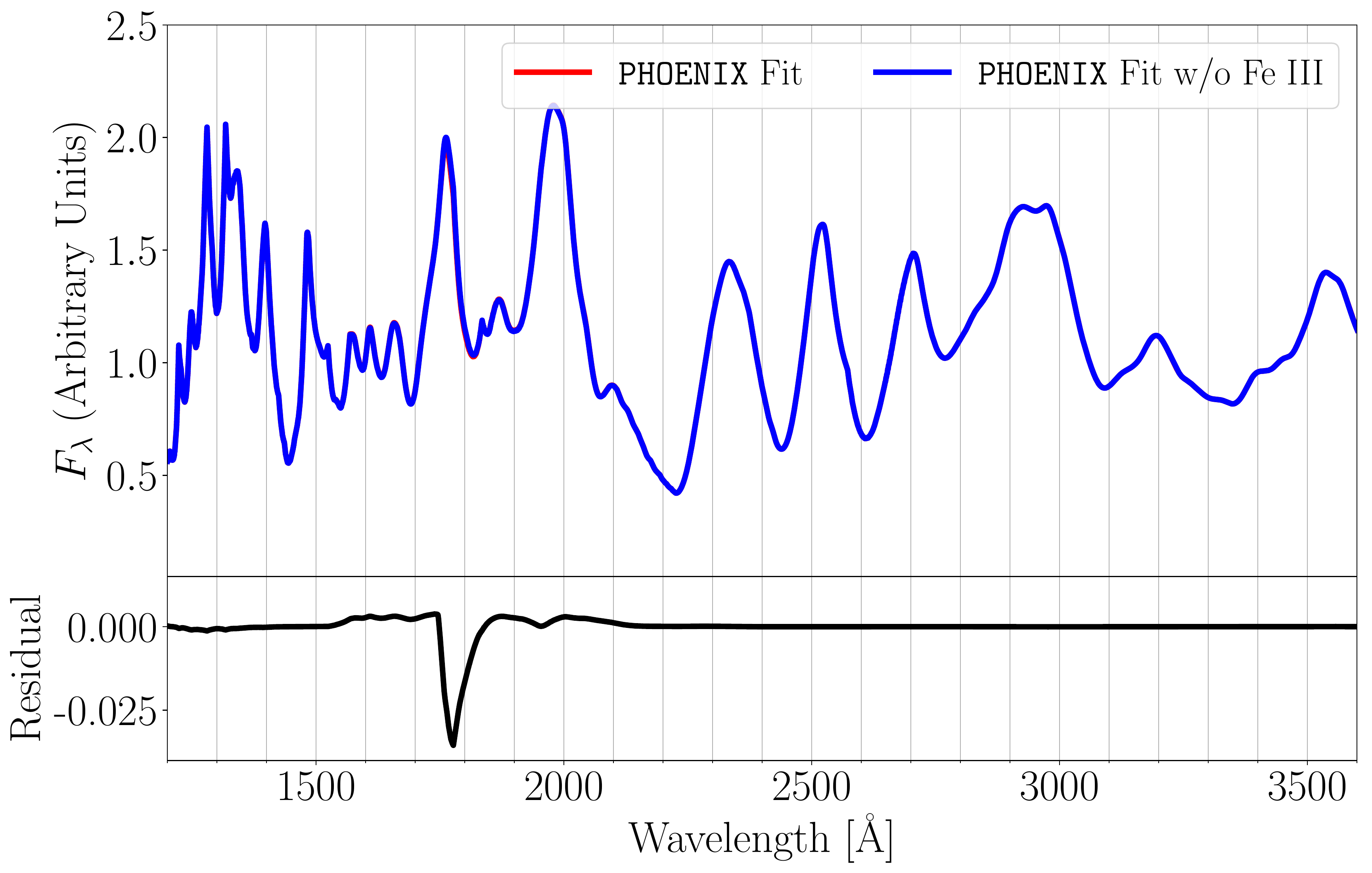}{0.45\textwidth}{(e) Fe III}
		  \fig{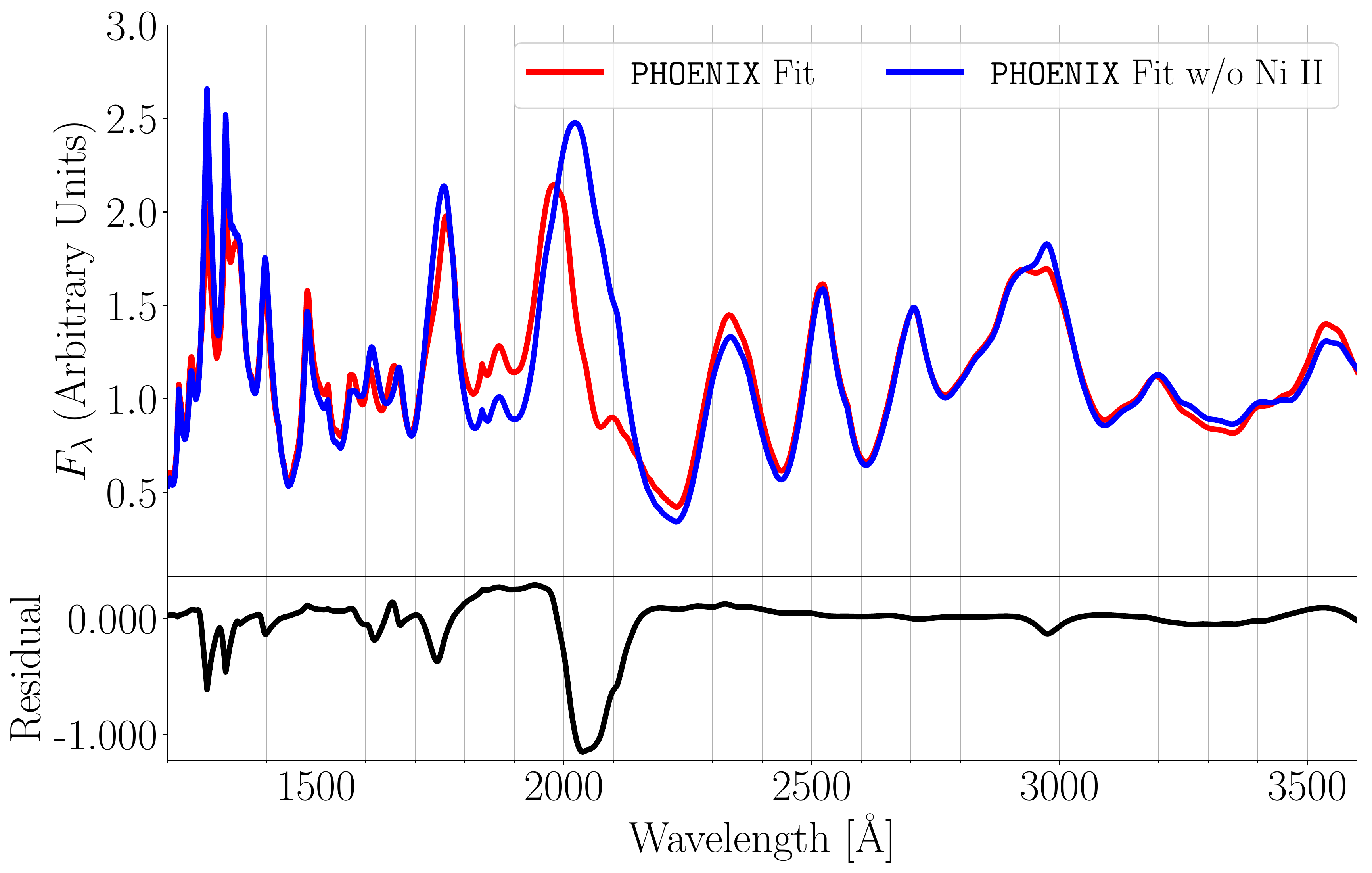}{0.45\textwidth}{(f) Ni II}}	  
\caption{Same as \cref{fig:ime_res} but for the iron group elements.}
\label{fig:ige_res}
\end{figure*}

\cref{fig:phx_id} shows the line identifications in the UV according the 
\texttt{PHOENIX} inverse single-ion spectra plotted in 
\cref{fig:ime_res,fig:ige_res}. Overall, the model replicates the observed spectrum 
well in the near and far-UV, but overestimates the flux levels in the mid-UV. 

Near-UV features are blends of IGEs including Fe~II and Co~II with the exception of the 
$\sim 2820$~\AA\ feature which includes a contribution from Mg~II ($\lambda\lambda2929,2937$).
In the mid-UV, features are similarly  due to Fe~II and Co~II, again with a contribution
from Mg~II ($\lambda\lambda2796,2803$) to the $\sim 2650$~\AA\ feature. There is also an 
isolated Ni~II feature at $\sim 2080$~\AA. The $\sim 2470$~\AA\ and $\sim 2650$~\AA\ 
features have the correct shape but are blueshifted relative to the observations.  
M14 also had difficulties in fitting sections of the UV spectrum with 
their WDD1 model, with the mid-UV fit showing a similar blueshift in the feature 
minima, albeit with flux values that were too low. This suggests that these 
blueshifts in the mid-UV feature minima are inherent in all delayed-detonation 
models, possibly due to the shape of the density profiles in the regions where these 
features form. The likely cause of the mid-UV flux differences are differing 
metalicities in the models ($0.5Z_{\odot}$ in M14 compared to $0.05Z_{\odot}$ 
here). This would support the conclusions of \citet{foley_kirshner} that 2011fe 
and 2011by differ in flux in the mid-UV due to differing metalicities. However, 
full exploration of the impacts of varying the metalicity in delayed-detonation 
models is beyond the scope of this work. 

In the far-UV, blends are much more complicated, with lines from Fe~II and Ni~II 
blending with a complex of Mg~II lines with rest wavelengths of $\sim 1480$~\AA\ 
and Si~II($\lambda\lambda1527,1533$) near the resonance lines from highly 
ionized species like C~IV ($\lambda\lambda1548,1551$) and Si~IV 
($\lambda\lambda1394,1403$). Determining the location of the C IV and Si IV 
resonance lines from the minimum of the residual yields velocities of $22,700$~km~s$^{-1}$ and $19,300$~km~s$^{-1}$. Similarly, the Si II and Mg II UV features 
are also formed in the outer layers of the ejecta with velocities of $21,600$~\kmps
and $18,000$~\kmps , $18,600$~\kmps, and $20,200$~\kmps 
for the $\sim 1480$~\AA\, $\lambda\lambda2796,2803$, and 
$\lambda\lambda2929,2937$ lines respectively. This indicates that the
UV resonance lines can be an important probe of the nature of the
outermost part of SNe Ia ejecta. Particularly the C~IV line can give
information on both the carbon abundance and the highest velocity of
the ejecta, providing clues to the SNe Ia environment.

To our knowledge C~IV and Si~IV have never before been identified in a SN. While
lines of singly and triply ionized C and Si are seen, very little, if
any doubly ionized lines are identified in the UV. Unlike the doubly ionized states, where the 
paired valence electrons increase the excitation energies and force the resonance 
lines into the unobservable UV, the triply ionized states have only one easily excited
valence electron, which allows the resonance lines to be located in the far-UV. All
other lines of the triply ionized state are high excitations and therefore unobserved.

\subsection{Agreement with SYNOW Line ID's}

In general, the line identifications from the SYNOW and \phx fits agree with each
other, with a few exceptions. In the near-UV, both fits agree that the observed 
features are due to blends mainly consisting of Fe~II and Co~II. However, the 
\texttt{PHOENIX} model disagrees that Cr~II is a significant contributor to the
$\sim 2790$~\AA\ and $\sim 3300$~\AA\ features. While a Cr~II line 
is present at the $\sim 3300$~\AA\ in the residual in \cref{fig:ige_res}, the 
line at this location is too weak to be considered a detection. This is perhaps due 
to the relatively small amount of Cr~present in the model, but may also be due to 
line blanketing or NLTE effects included in \phx but not SYNOW. Both synthetic 
spectra find that the mid-UV can be modeled by blends of Mg~II, Cr~II, Fe~II, 
and Co~II, although there is some disagreement over the specific compositions 
of individual blends (ex. Fe~II and Co~II in the $\sim 2470$~\AA\ and 
$\sim 2250$~\AA\ features). Both fits also find support for an isolated 
Ni~II feature at $\sim 2080$~\AA. In the far-UV, both fits agree on
the necessity of C~IV and Si~IV to fit the observed features, however
they disagree slightly on the  
weaker components of the blends, with the \texttt{PHOENIX} fit showing evidence 
of Si~II, Mg~II and Ni~II in the $\sim 1430$~\AA\ blend, while not finding 
significant Co~II in the $\sim 1290$~\AA\ feature. Throughout the UV, 
both the SYNOW and \phx fits find doubly ionized iron group species make no 
significant contributions to any features.

\subsection{Model Agreement in Optical and NIR}

\begin{figure*}[ht]
\centering
\gridline{\fig{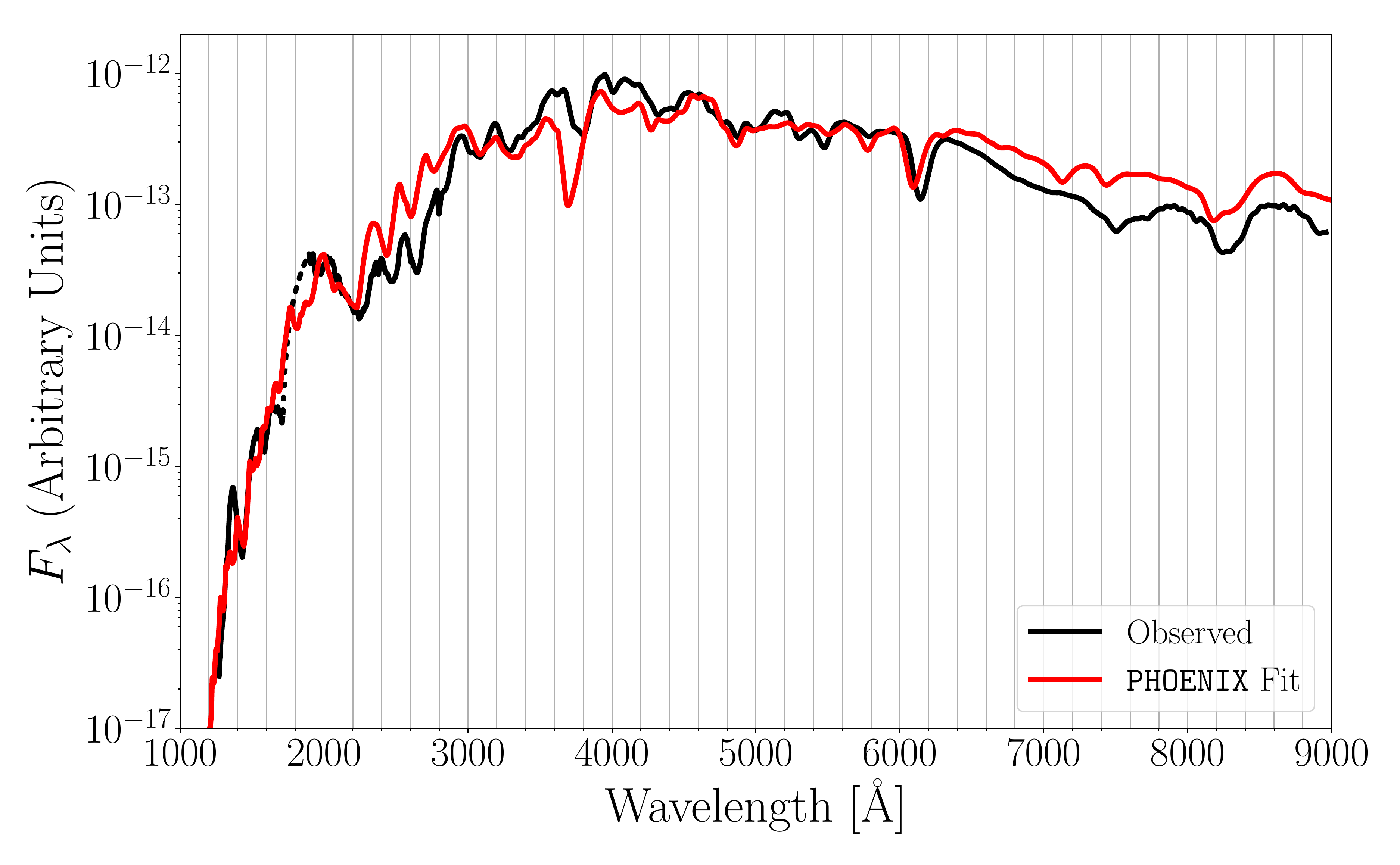}{0.45\textwidth}{(a)}
		  \fig{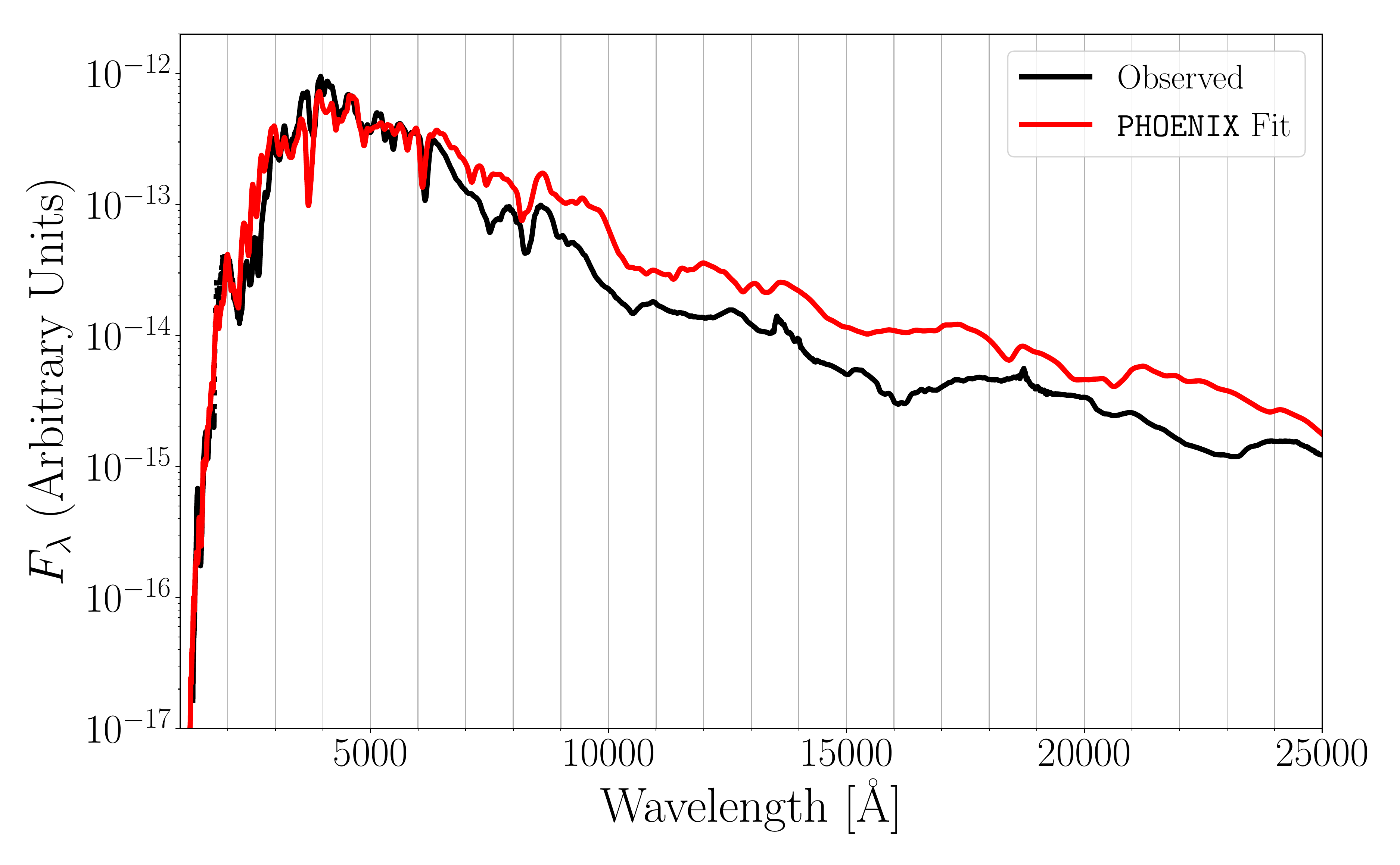}{0.45\textwidth}{(b)}}
\caption{The best fit \texttt{PHOENIX} spectrum at day 23 (in red) compared to 
observations (black) spanning from the far-UV to near-IR. The target luminosity 
of the model is $2.76 \times 10^{42}$ erg s$^{-1}$. By including the UV in addition 
to the optical in the determination of the best fit spectrum, the optical fit is 
marginally worse than models where only the optical spectrum is considered in 
determining the best fit. Differences can primarily be seen in the flux levels 
near the Ca II H \& K and NIR-triplet features.} 
\label{fig:phx_full_L6}
\end{figure*}

An important test for the validity of delayed-detonation models is the ability
to simultaneously reproduce observed spectra across a large wavelength range.
\cref{fig:phx_full_L6} shows the best fit model is able to match the observed
spectrum not only in the UV, but the optical and NIR as well. While the model is
able to reproduce the spectrum in the UV and optical, there is a consistent flux
excess in the NIR, extending to 2.5 microns. Using the same model, \citet{baron2015}
were able to replicate the optical and NIR spectra (see their Fig. 11) at max
light but did not attempt to simultaneously fit the UV, which suggests the underlying model 
is limited in its ability to recreate the spectra in all three regions at once. 
This particular model seems to produce the Ca H+K line
  significantly stronger than observed, which is likely due to the
  structure of the explosion model itself.
Potential causes of these flux mismatches are further explored in
Section \ref{section:temps}.

\section{Other Spectral Formation Mechanisms} \label{sec:spec_form}

\subsection{Photoionization Edges}

\begin{figure*}
\centering
\gridline{\fig{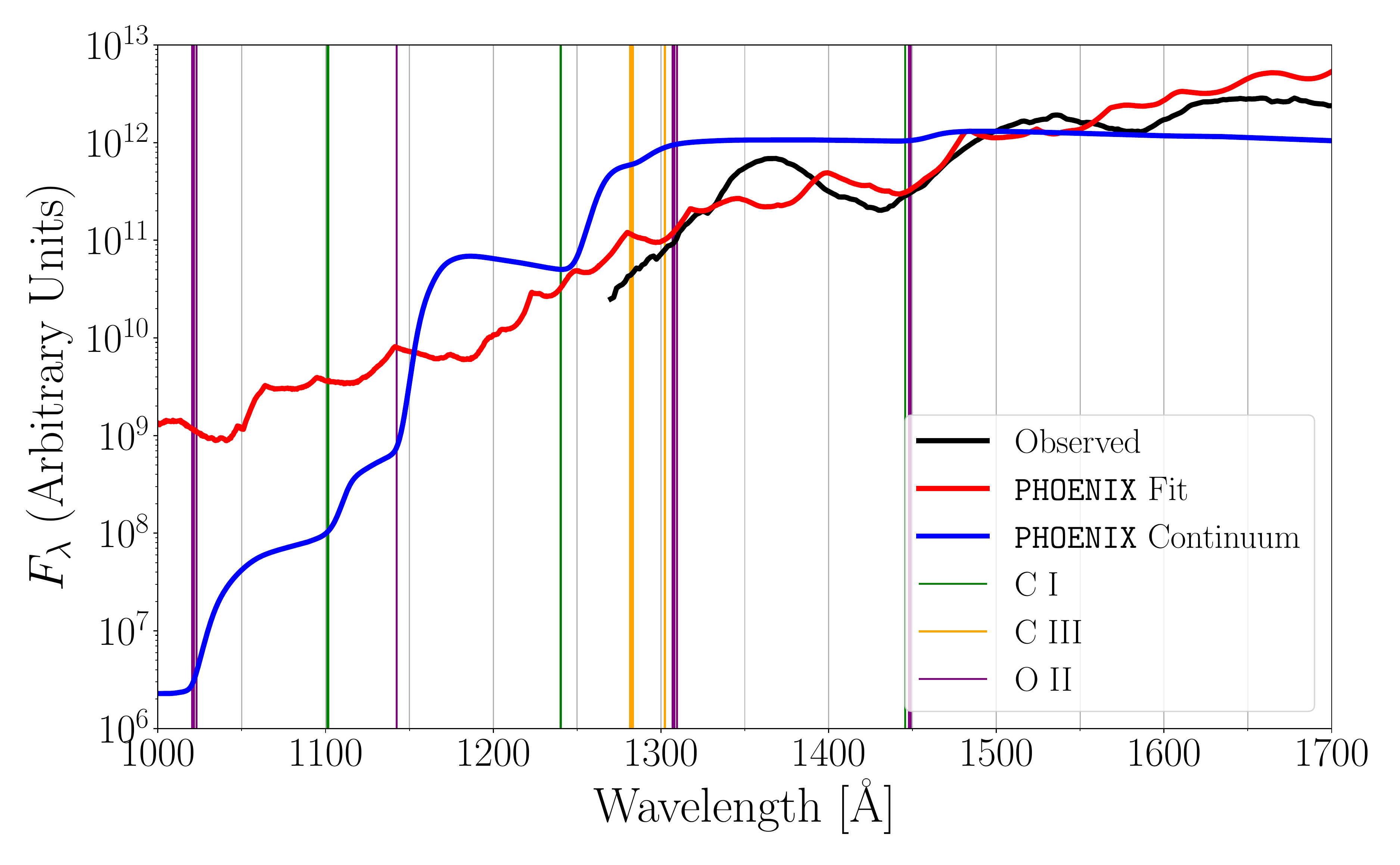}{0.45\textwidth}{(a)}
		  \fig{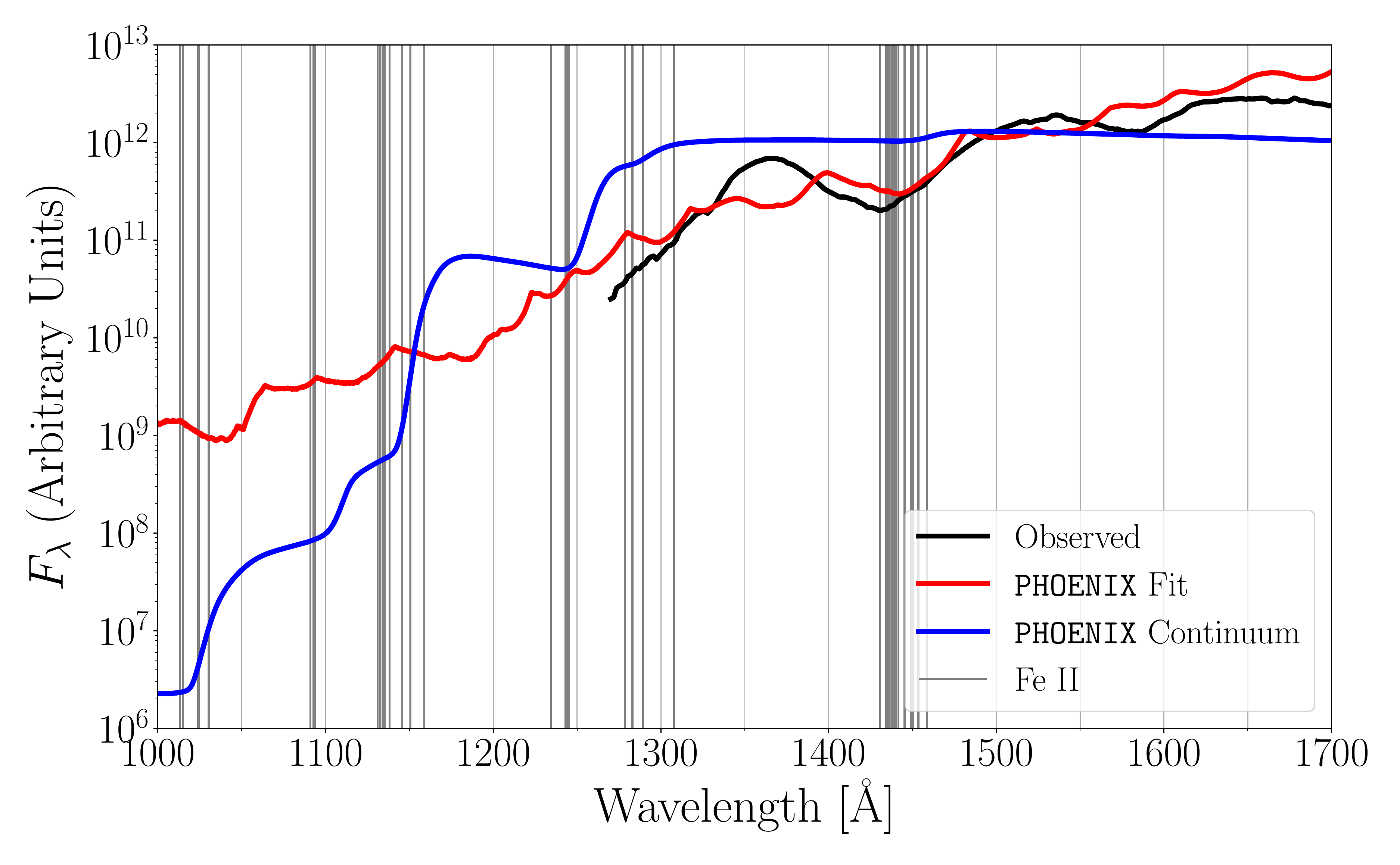}{0.45\textwidth}{(b)}}
\caption{Full \phx fit (red), compared to observations (black) and the
continuum-only fit (blue) scaled to the observed flux at 1500 \AA. 
Photoionization edges due to C~I (green), C~III (orange), and O~II (purple)
are noted with vertical lines in panel (a). Panel (b) shows the same plot, with
only Fe~II photoionization edge locations in grey. The plot has been extended 
to show additional photoionization edges imprinted on the continuum fit beyond 
the wavelength range of the observed spectrum. The photoionization edges which 
are not easily visible in the full fit, are coincident with and contribute to 
the feature formation in the far-UV.}
\label{fig:ion_edges}
\end{figure*}

At the threshold wavelength for a bound-free transition, the opacity
jumps, since redward of that wavelength the opacity in that transition
is zero, and as one reaches the threshold energy, the opacity is
finite. This jump in opacity is not dissimilar to what happens in a
line, and thus at the threshold wavelength of bound-free
(photoionization edges), one sees P-Cygni like features, with only
continuum opacity and all line transitions ignored
\citep{baron99_94I}.
These photoionization edges are easily seen in ``continuum-only 
spectra", which are generated in a manner similar to ``single-ion spectra" 
but with the opacities of all lines artificially set to zero, leaving only the 
bound-free and free-free processes to determine the opacity. The continuum-only 
spectra of \citet{bongard2008} show in the W7 model near maximum light several 
photoionization edges form blue-wards of 2000~\AA\ (see their Figure 5). 

Examining the ionization energies of elements present in the model reveals two
potential sources of the edges. The first is ionization of C and O from low 
excitation states in the outer layers of the ejecta. The second is ionization 
from higher excitation states of Fe~II located deep within the ejecta. 
We create continuum-only spectra of our DD model, shown in \cref{fig:ion_edges} 
with panel (a) showing the wavelengths of ionization edges from C~I, C~III, and O~II; 
while panel (b) shows the ionization locations of Fe~II. All photoionization edges in 
our model form in the far-UV, and some are coincident with the features that form 
at $\sim 1430$~\AA\ and $\sim 1290$~\AA. In both cases, not all excited
states correspond to an observed edge in the model, and several edges seen in the model
may in fact be combinations of edges from different ions (ex. the edge at $\sim 1450$~\AA\ 
is located near ionizations of C~I, O~II, and Fe~II).

\subsection{Line Blanketing}

\begin{figure}
\centering
\includegraphics[width=0.9\columnwidth]{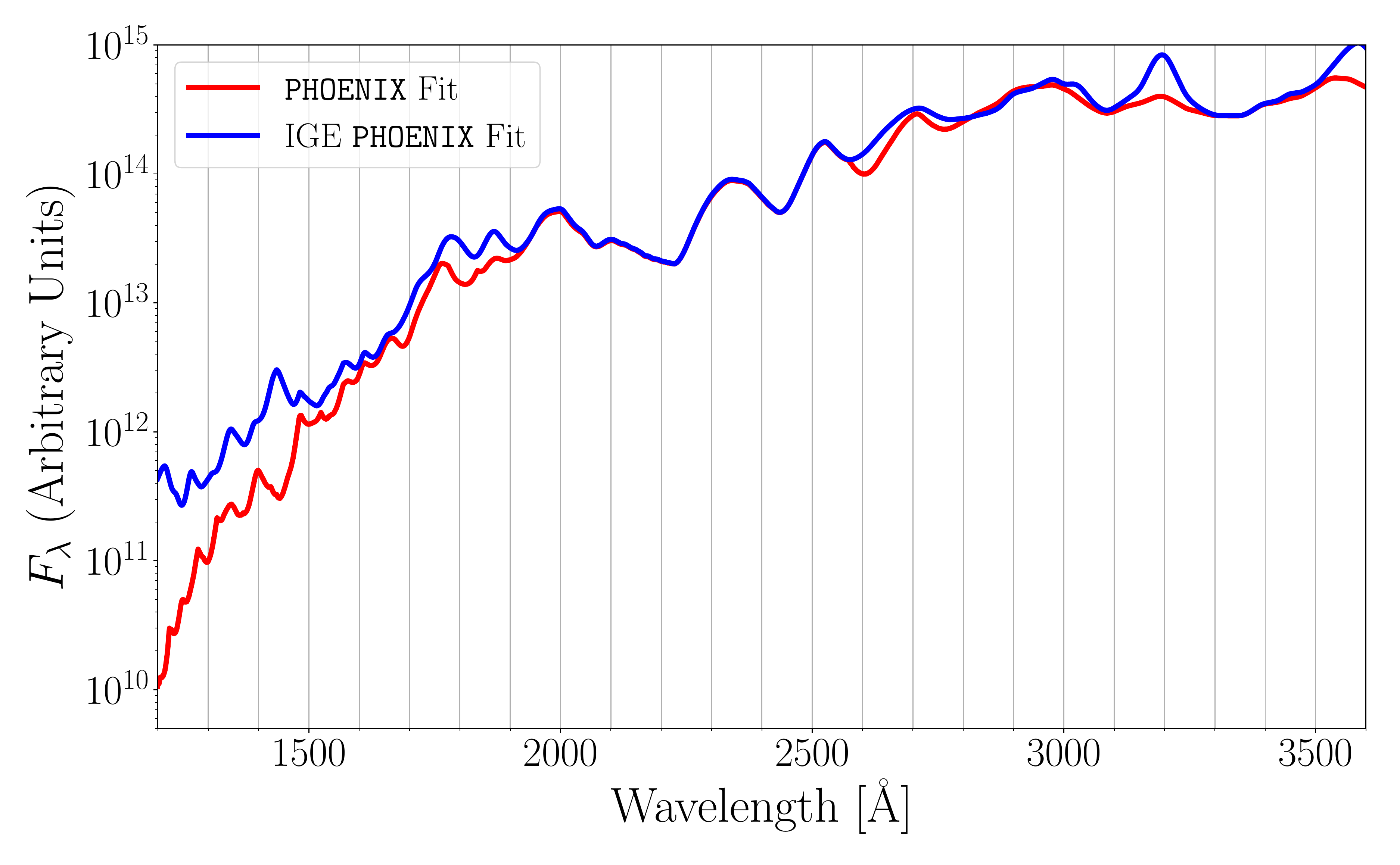}
\caption{The full fit \texttt{PHOENIX} spectrum (red), compared to the best fit 
model with only IGE (Cr~I-III, Mn I-III, Fe~I-III, Co~I-III, Ni~I-III) lines 
included (blue). }
\label{fig:line_blanketing}
\end{figure}

Line blanketing from IGE's has long been thought a dominant factor in determining
the flux levels in the UV spectra of Type Ia SNe. Utilizing \texttt{PHOENIX}'s 
ability to artificially remove elements from the spectra by setting the opacities of
specific elements to zero, we create an IGE-only spectrum, containing only the lines
of Cr~I-III, Mn I-III, Fe~I-III, Co~I-III, and Ni~I-III to determine the importance
of this effect. The IGE-only spectrum is shown in \cref{fig:line_blanketing} compared
to the best fit spectrum. In the near and mid-UV, the IGE only-spectrum is able to 
reproduce almost exactly the observed features and flux levels, excluding the 
contribution of Mg~II in the mid-UV and the emission peak around 3200 \AA. The 
tight agreement of the IGE-only spectrum with that of the full fit suggests that
the IGE's are the primary  drivers of  spectral formation in this region, and
measurements of this region can provide insight into quantities
related to the velocity extent of
iron group elements as well as their abundances. 

However, in the far-UV the flux of the IGE-only fit slowly deviates from that of
the full fit until it is an order of magnitude too high, indicating that lines from 
unburned and partially burned material in the outermost layers of the ejecta play an 
important role in the formation of this region of the spectrum. Further insights into 
this material may be gained from additional observations of Type Ia spectra in the far-UV.  

\subsection{Temperature Dependences} \label{section:temps}

Several spectral features, both in the optical and ultraviolet
demonstrate strong temperature dependence in our  
models. We briefly describe a few of these 
ultraviolet features here, while a more complete analysis of all temperature
dependent features in our models from ultraviolet to near-infrared is
left to future work (J.~DerKacy, et al., in prep).
We should note that we use temperature in a very general sense
  to denote variations in the spectral energy distribution and
  ionization state throughout the atmosphere. Our models, are highly
  NLTE and thus one cannot capture either the state of the radiation
  field or the ionization state of the ions with one simple quantity, temperature.

\begin{figure}
\centering
\fig{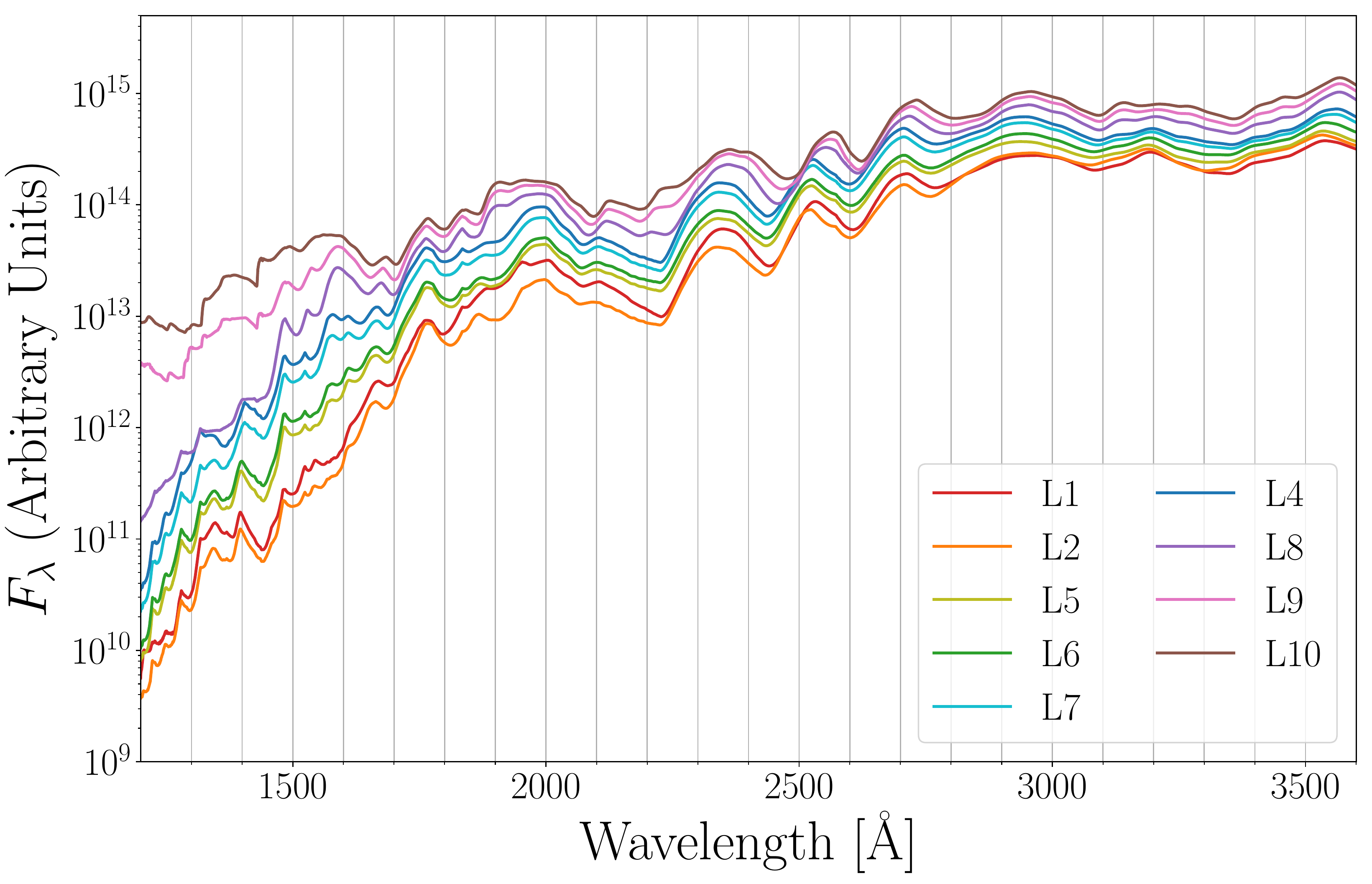}{0.9\columnwidth}{}
\caption{The suite of \texttt{PHOENIX} models with different target luminosities 
showing the evolution of the UV spectrum with temperature. The L6 model is the
best fit model. The target luminosities of the models (in erg s$^{-1}$) are - L1: 
$7.50 \times 10^{42}$, L2: $7.90 \times 10^{42}$, L5: $8.29 \times 10^{42}$,
L6: $8.69 \times 10^{42}$, L7: $9.08 \times 10^{42}$, L4: $9.47 \times 10^{42}$, 
L8: $1.03 \times 10^{43}$, L9: $1.11 \times 10^{43}$, and L10: $1.18
\times 10^{43}$. The temperatures at the electron scattering
  optical depth $\tstd = 2/3$ are L1: $12.1 \times 10^3$~K, L2:
  $12.2 \times 10^3$~K, L5: $12.0 \times 10^3$~K, L6: $12.3 \times 10^3$~K,
  L7: $12.2 \times 10^3$~K, L4: $12.2 \times 10^3$~K, L8: $12.1 \times 10^3$~K,
  L9: $12.3 \times 10^3$~K, L10: $12.4 \times 10^3$~K. Due to variations in the
  ionization state which causes variations in the position of $\tstd = 2/3$ and
  the NLTE nature of the ionization states, the temperatures are not perfectly
  monotonic, showing the extent of the deviation of the models from LTE.}
\label{fig:uv_models}
\end{figure}

To better understand this temperature dependence we created several runs of our
underlying DD model, each with a different target luminosity. The underlying 
density and abundance structure was held constant in each run. Changing the target 
luminosity between runs alters the temperature structure, opacities, and ionization 
balance of the runs, while they are iterated to radiative equilibrium. The most temperature 
sensitive regions of the spectra are the far-UV, where variations are seen in both 
the overall flux level and the C~IV blend near 1430~\AA, and the mid-UV Fe~II blends 
near $\sim 2470$~\AA\ and $\sim 2750$~\AA, and the full UV spectrum 
are shown in \cref{fig:uv_models}.

\begin{figure}
\centering
\fig{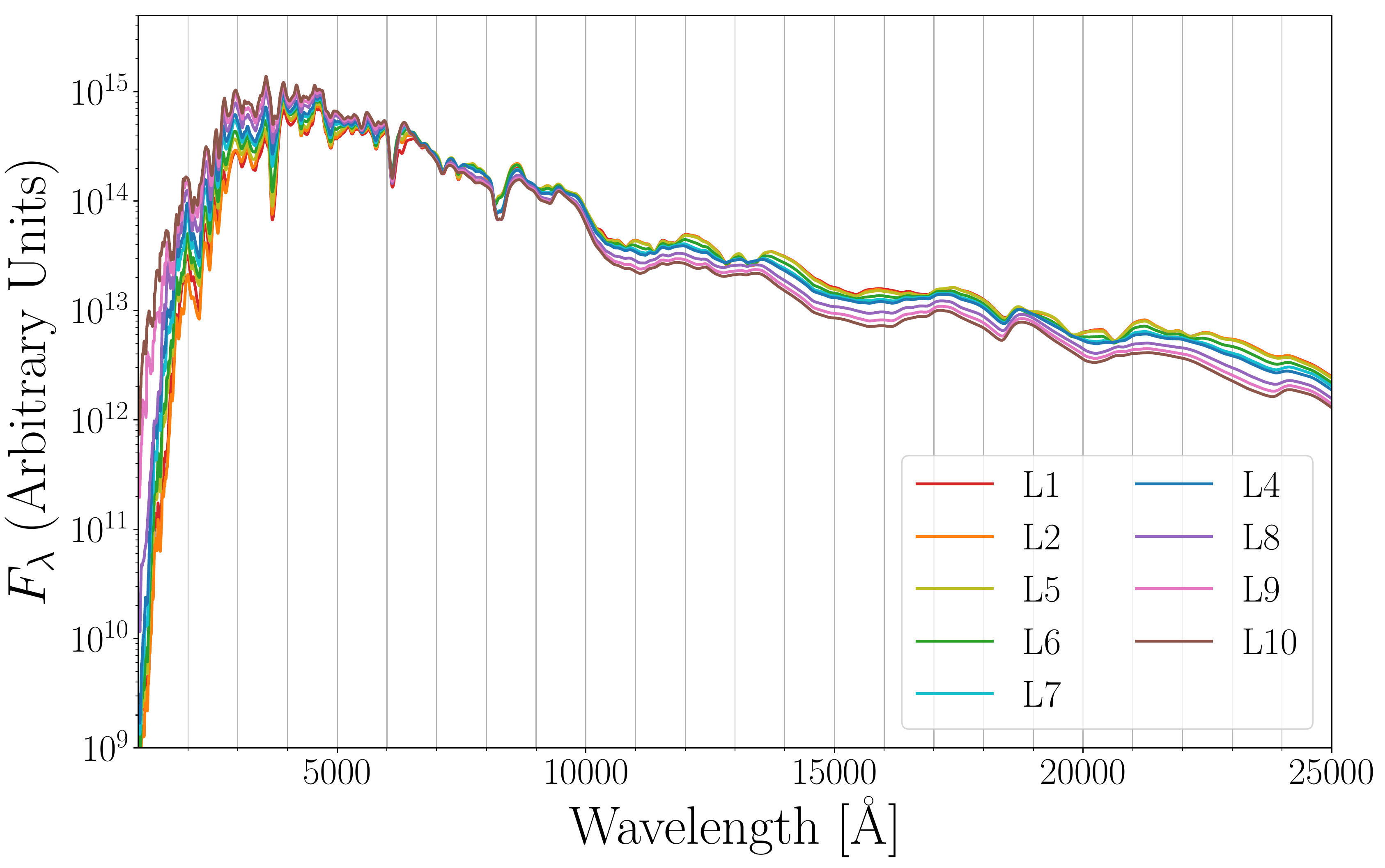}{0.9\columnwidth}{}
\caption{Same as \cref{fig:uv_models} but showing NIR instead.}
\label{fig:nir_temps}
\end{figure}

\begin{figure}
\centering
\fig{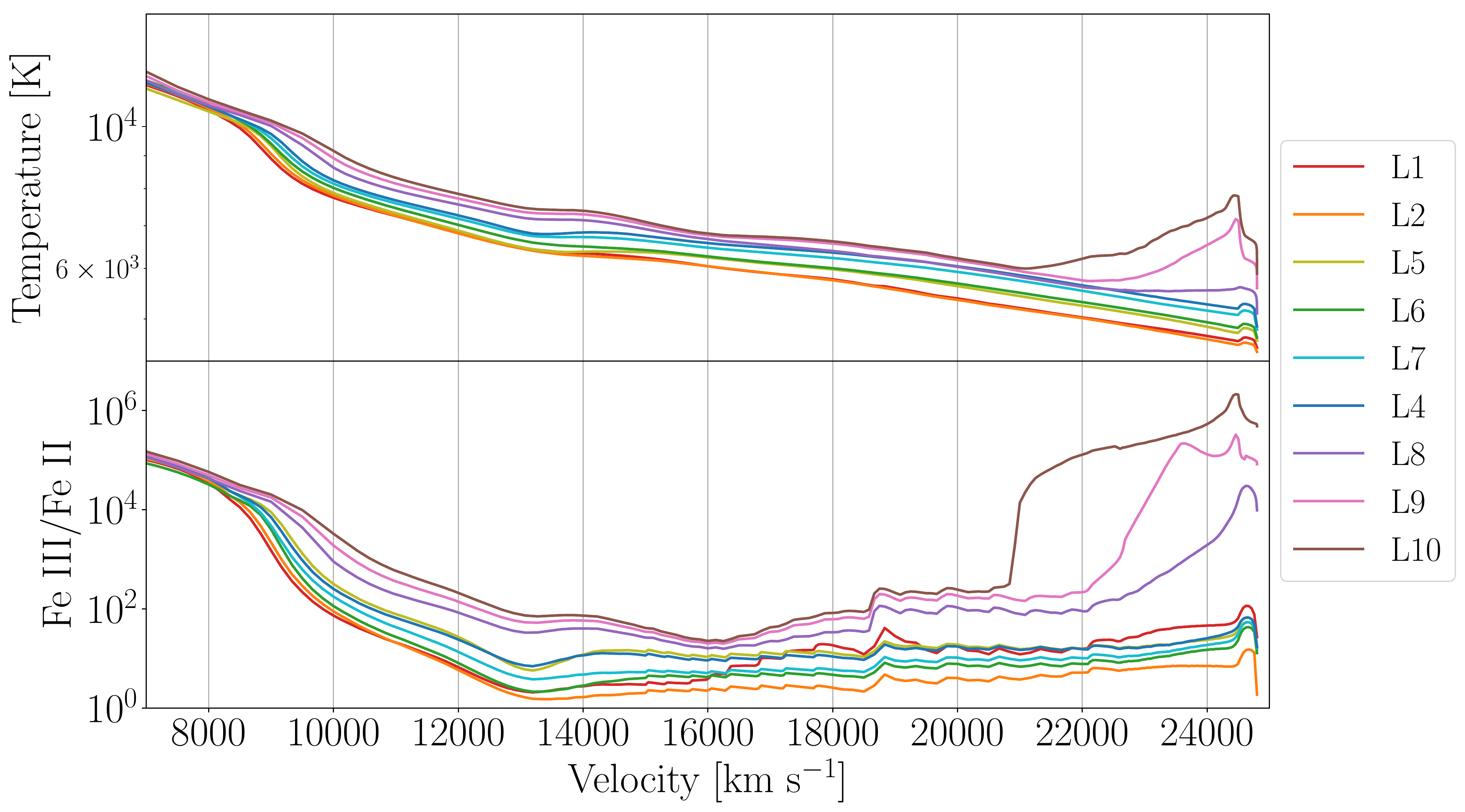}{0.9\columnwidth}{}
\caption{The top panel shows the temperature structure of the various models,
while the bottom panel shows the ratio of Fe~III/Fe~II. Below 7000 km s$^{-1}$
all the models have nearly identical temperature structures and ionization balances.}
\label{fig:fe_ratio}
\end{figure}

The most noticeable difference in the spectra appears in the far-UV where the flux level
of the spectra varies by three orders of magnitude across the runs. This drastic 
change in flux is caused by the changing ratio of Fe~III/Fe~II above the photosphere, 
as seen in \cref{fig:fe_ratio}. Because Fe~II is more efficient at redistributing flux 
from blue to red than Fe~III, runs with lower temperatures that have a higher fraction 
of Fe~II in the outer layers have lower UV flux levels and increased NIR fluxes. The 
variations in the NIR spectra can be seen in \cref{fig:nir_temps}. As the temperature 
in the outer layers is increased, the the UV (and especially far-UV) 
flux begins to deviate more from the observed spectrum while the NIR flux levels slowly
come into agreement with observations. 

\section{Discussion} \label{sec:discussion}

The ability of the model to match either the UV or NIR spectra, but not both, in
addition to the optical spectra suggests several potential changes that may help to
bring the model into better agreement with the observations. One potential change 
is to adjust the model's metalicity. This adjustment may prove difficult however, as 
any potential decrease in UV flux due to increased metalicity must be balanced against 
any increases in flux that result from higher opacities pushing the photosphere farther 
out in the ejecta and higher temperatures in the outer layers altering 
the ionization balance of the outer layers; in particular the Fe
III/Fe~II ratio.  This is the subject of future work, where we will
self-consistently study the metalicity dependence and compare to a
broad range of observed SNe Ia.

Secondly, an incorrect distribution of  $^{56}$Ni could also play a
role in the flux mismatch. Broad wavelength coverage can provide clues
as to the velocity extent of \nni
\citep{Ashall_Hband19a,Ashall_Hband19b}. 
From the SN~2011fe light curve \citet{piro11fe12} 
concluded that \nni was required in
the outer $0.1 < M < 1 \times 10^{-3}$ of the supernova, however,
\citet{baron2015} found that \nni at such high velocities did not affect
the spectra, but they were focused on the optical and infrared and not
the UV. 

Finally, the density structure of the model, which beyond
18,000~km~s$^{-1}$ closely approximates an $n=7$ power-law, 
may not accurately describe the density profile in the outer layers. \citet{sauer2008}
were able to produce similar results in their simulations of W7 by varying the power law
exponent above 15000 km s$^{-1}$. Models with steeper density gradients produced lower
flux values in the far-UV as they had a lower Fe~III/Fe~II ratio. However, the increased
density in the outer layers result in smoother, almost featureless spectra. Models with
shallower density gradients produced higher flux values due to high Fe~III/Fe~II ratios and
appropriate spectral features. Taken together with our results, this suggests that temperature
and density variations in the outer ejecta layers are degenerate with respect to the far-UV spectra.
However, we stress that the full investigations of these effects needed to break this 
degeneracy are beyond the scope of this work. 

\begin{figure*}
\centering
\gridline{\fig{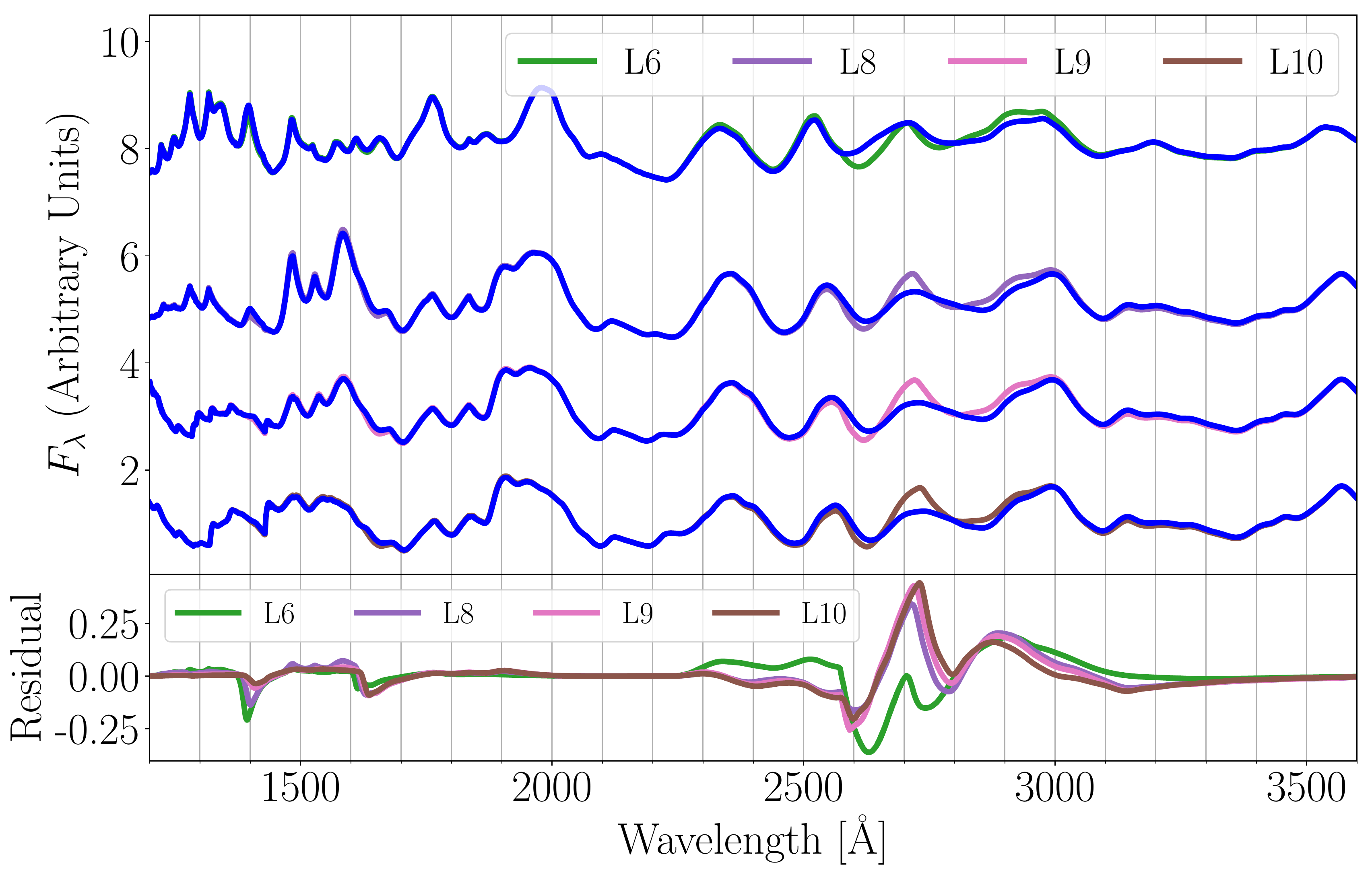}{0.45\textwidth}{(a) Mg II}
		  \fig{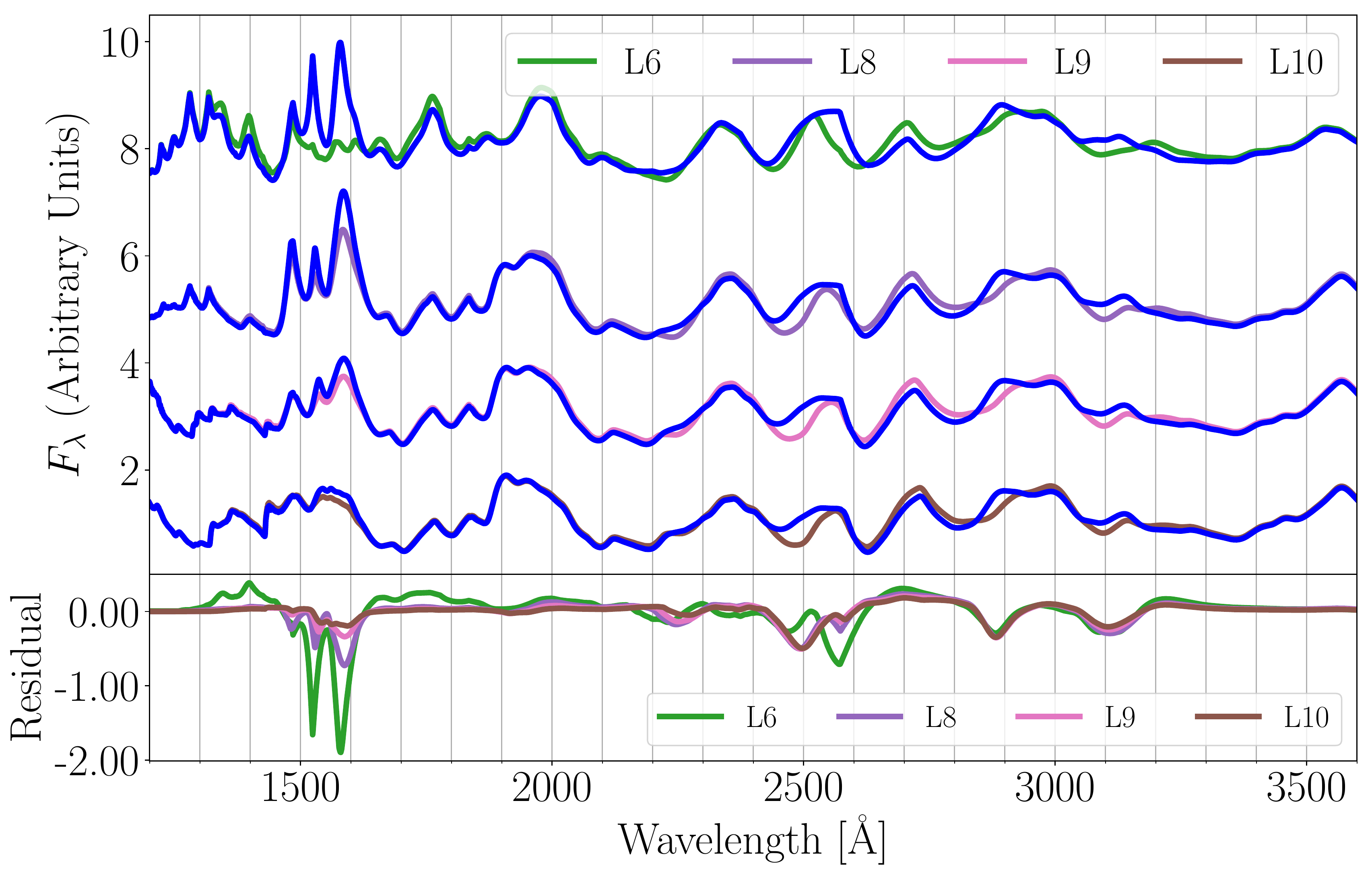}{0.45\textwidth}{(b) Fe II}}
\caption{Residual plots of Mg~II and Fe~II as in \cref{fig:ime_res,fig:ige_res} for the
L6 (green), L8 (purple), L9 (pink), and L10 (brown) models. The blue line overlaying 
each model fit is the inverse single-ion spectrum for Mg~II (panel a) or
Fe~II (panel b) for that run luminosity. Residuals between the model and inverse 
single-ion spectra share the same coloring as the model in the 
top panel.}
\label{fig:res_temp}
\end{figure*}

The variation in the Fe~III/Fe~II ratio also partially explains the velocity 
shift in the Fe~II blends in the mid-UV. As the amount of Fe~II at higher velocities 
decreases with the increased temperature, the Fe~II lines form deeper in the ejecta 
at lower velocities, bringing the model into better agreement with the observed 
velocities of the feature. This change in the wavelength of the Fe~II feature forming 
the blend at $\sim 2650$~\AA\ could also be the result of the higher temperature 
favoring a different combination of Fe~II lines in the region. Additionally, the higher 
temperatures drive an increase in the strength and width of the emission peak of the 
Mg~II feature at $\sim 2650$~\AA\ due to $\lambda\lambda 2796,2803$. This in turn 
causes the absorption minima of the $\sim 2820$~\AA\ feature to appear redder, in
better agreement with the observations. \cref{fig:res_temp} shows the evolution of the Fe~II
and Mg~II residuals with increasing temperature.

\section{Conclusions} \label{sec:conclusion}

We fit the UV spectrum SN 2011fe at +3.4 days after maximum light with synthetic
spectra generated by both SYNOW and \texttt{PHOENIX} (with a DD model) to provide line 
identifications for all major features. Both codes generally agree in their 
identifications, with both the near and mid-UV spectra comprised of blends of 
Cr~II, Fe~II, Co~II and Ni~II. Features in the far-UV are formed by strong 
resonance lines of C~IV, Si~II, and Si~IV combined with the now less dominant 
lines from singly ionized IGE's that dominate the rest of the UV spectra. 
Leveraging the ability of \phx to generate ``single-ion spectra", we 
further investigate the impact of other spectral formation mechanisms on the UV 
spectra. We find several photoionization edges coincident with the 
far-UV features which significantly contribute to determining the flux level in 
this region. We also examine an IGE-only spectrum, which is able to reproduce
both the features and flux levels of a significant portion of the UV spectrum,
further confirming the utility of mid-UV measurements as a probe of IGEs in Type Ia SNe.
Finally, using a suite of \texttt{PHOENIX} models with different
target luminosities, we identify several regions of the UV spectrum with strong
temperature dependence. In the far-UV, the Fe~III/Fe~II ratio in the outermost
portion of the ejecta changes rapidly with temperature and results in a flux
levels that vary by almost four orders of magnitude across the models. In the
mid-UV, the features centered at $2250$~\AA\ and $2470$~\AA, which
are too fast in DD models, begin to recede in velocity and better match observations.

DD models both here and elsewhere have shown the ability to replicate the spectra of
Type Ia SNe in the UV and future work using DD models which implicitly account for 
metalicity and density variations are needed to investigate the UV diversity of Type 
Ia SNe in greater detail.

\acknowledgments

E.B. and J.D. are supported in part by NASA grants NNX16AB25G, 
NNX17AG24G, and 80NSSC20K0538.
Some of the  calculations presented here were performed at the
H\"ochstleistungs Rechenzentrum Nord (HLRN), 
at the National Energy Research Supercomputer Center (NERSC), which is
supported by the Office of Science of the U.S.  Department of Energy under
Contract No. DE-AC03-76SF00098 and at
the OU Supercomputing Center for Education \& 
Research (OSCER) at the University of Oklahoma (OU).
We thank all these institutions for a generous
allocation of computer time.

\software{SYNOW \citep{jeffery_branch_1990,fisher2000}, \texttt{PHOENIX} \citep{phx1999}}

\bibliography{11fe_UV_bib}

\end{document}